\documentclass[12pt,english]{article}
\usepackage{
amsmath,
amssymb,
babel,
braket,
booktabs,
caption,
cite,
float,
graphbox,
graphicx,
hyperref,
pgfplots,
siunitx,
subfig,
subfloat,
tabu,
xcolor,
xparse,
ytableau
}
\pgfplotsset{compat=1.13}
\makeatletter

\newcommand{\eqb}{\begin{equation}}
\newcommand{\eqe}{\end{equation}}
\newcommand{\dmb}{\begin{displaymath}}
\newcommand{\dme}{\end{displaymath}}

\newcommand{\eab}{\begin{eqnarray}}
\newcommand{\eae}{\end{eqnarray}}

\newcommand{\be}{\begin{equation}}
\newcommand{\ee}{\end{equation}}
\newcommand{\sgn}{\text{sgn}\,}

\hypersetup{colorlinks=true,
	linkcolor={red!30!black},
    	citecolor={blue!50!black},
    	urlcolor={blue!80!black}
   	 }

\RenewDocumentCommand\[{}{\begin{equation}}
\RenewDocumentCommand\]{}{\end{equation}}
\NewDocumentCommand\der{}{\mathrm{d}}

\NewDocumentCommand\angl{}{\sphericalangle}
\NewDocumentCommand\gsim{}{\gtrsim}

\NewDocumentCommand\epsi{m}{\epsilon_{#1}}

\NewDocumentCommand\intkern{}{\int\kern-5pt}
\NewDocumentCommand\mat{m}{\left(\begin{matrix}#1\end{matrix}\right)}

\makeatother

\begin{document}
\begin{titlepage} 

\vspace{0.6cm}

\begin{center}
{\Large {Massive loops in thermal SU(2) Yang-Mills theory:\\ Radiative corrections to the 
pressure beyond two loops} \vspace{1.5cm}
}\\
{\Large{} }{\large {Ingolf Bischer$^\dagger$, Thierry Grandou$^*$, and Ralf Hofmann$^\dagger$} }
\par\end{center}
\vspace{1cm}
\begin{center}
$^*$ Universit\'{e} Cote d' Azur,\\ Institut de Physique de Nice,\\ Routes des Lucioles, 06560 Valbonne, France
\end{center}
\vspace{1cm}
\begin{center}
$^\dagger$ Institut f\"ur Theoretische Physik, \\
Universit\"at Heidelberg, \\
Philosophenweg 16, 69120 Heidelberg, Germany
\end{center}

\vspace{1cm}
 
\begin{abstract}
We device an efficient book-keeping of excluded energy-sign and scattering-channel combinations 
for the loop four-momenta associated with massive quasi-particles, circulating in (connected) 
bubble diagrams subject to vertex constraints inherited from the thermal ground state. 
The according radiative corrections modify the free thermal-quasiparticle pressure 
at one loop. Increasing the loop order in two-particle irreducible (2PI) bubble diagrams, we exemplarily demonstrate a 
suppressing effect of the vertex constraints on the number of valid combinations. This increasingly strong 
suppression gave rise to the conjecture in hep-th/0609033 that the loop expansion would terminate at a finite order. 
Albeit the low-temperature dependence of the 2PI 3-loop diagram complies with this behaviour,  
a thorough analysis of the high-temperature situation reveals that 
the leading power in temperature is thirteen such that this diagram
 dominates all lower loop orders for sufficiently high temperatures. An all-loop-order 
resummation of 2PI diagrams with dihedral symmetry is thus required, defining an extremely 
well-bounded analytical continuation of the low-temperatures result.   
\end{abstract}
\end{titlepage}

\tableofcontents
\section{\label{SEC:1}Introduction}
Four dimensional Yang-Mills thermodynamics is a fascinating subject 
\cite{Hofmann2016}. However, attempts to tackle the high-temperature 
Yang-Mills problem in terms of small-coupling expansions \cite{thermalPT1,thermalPT2,thermalPT3,thermalPT4,thermalPT5,thermalPT6,thermalPT7} lead  
to a poorly controlled foliation \cite{Linde1980} of (likely infinitely many) 
hierarchical momentum scales \cite{thermalPT8,BGHwip}. The integration of this 
hierarchy presumably represents an unsurmountable technical difficulty 
and conceptually {\sl is} questionable. The problems induced by the  
perturbative approach can be traced to the disregard of an appropriate 
deconfining thermal ground-state estimate which is composed of Yang-Mills field 
configurations that, as a matter of principle, are inaccessible 
to small-coupling expansion: Harrington-Shepard (anti)calorons 
\cite{HS1977} for SU(2) and SU(3) gauge groups 
\cite{Hofmann2016}. In fact, propagating disturbances and quantum fluctuations as well 
as their residual correlations are direct consequences of the excitability of this ground state 
\cite{Entropy2016}. Since already on the level 
of free excitations a part of the spectrum is represented by thermal 
quasiparticles due to an adjoint Higgs mechanism, since the ground 
state is associated with a scale of maximum resolution, and 
since their is a straightforward physical and completely fixed gauge, 
radiative corrections are void of infrared divergences, and the ultraviolet 
cutoff can be prescribed in a physically meaningful way. 
Recall that thermal perturbation theory requires a 
renormalisation programme which performs subtractions in certain, technically convenient gauges 
by appealing to zero-temperature renormalisation conditions \cite{Evans,Linde1980}.  

The usefulness of a deconfining thermal ground state as an 
a priori estimate would be reflected by the fact that radiative 
corrections to free quasiparticle excitations are in some 
sense hierarchically small. Therefore, it is of utmost importance 
to construct an accordingly exhaustive expansion scheme. A seemingly natural expansion of radiative corrections 
is in loop orders \cite{Brodsky}. Note that such an expansion does not 
necessarily rely on the smallness of the coupling constant. In fact, the gauge coupling 
constant $e$ of deconfining SU(2) Yang-Mills thermodynamics is $e=\sqrt{8}\pi>1$ for almost 
all possible temperatures \cite{Hofmann2016}. 
In \cite{SGH2007} all two-loop corrections to the pressure were 
computed, and, compared to the one-loop result, a hierarchical suppression 
was observed\footnote{The three-loop results in \cite{KavianiHofmann} are 
erroneous due to the appeal to incorrect vertex constraints.}. 

In the present work we investigate whether such a hierarchical 
suppression with increasing but fixed loop order holds beyond 
two loops. To keep the technical complexity at a minimum, 
we consider loop diagrams that exclusively involve massive 
quasiparticles since these are characterised by 
thermal fluctuations only \cite{Hofmann2016}. In spite of the 
fact that the ratio of allowed energy-sign and scattering-channel combinations 
to such a priori possible combinations decreases with increasing loop order, we find 
that, starting at the 2PI three-loop order, the hierarchical 
suppression is spoiled for temperatures $T$ sufficiently far above the critical 
temperature $T_c$ of the deconfining-preconfining 
phase transition. This calls for a resummation within a certain 
class of all-order 2PI loop diagrams. Indeed, we demonstrate that such a 
resummation generates an analytic temperature dependence which is well 
bounded for all temperatures $T\ge T_c$.  

This paper is organised as follows. In Sec.\,\ref{SEC:2} 
we review the constraints on invariant momentum transfer through a 4-vertex involving massive modes 
only and, based on this, set up a counting scheme for excluded energy-sign and scattering-channel combinations. This scheme is subsequently applied to 2PI bubble diagrams up to six-loop order as they occur in the loop expansion of the pressure. We observe that the ratio $R$ of allowed vs. a priori possible combinations monotonically decreases with increasing loop number. However, we also 
prove that within a class of dihedrally symmetric diagrams $R$ never vanishes at 
any finite loop order and that therefore a more quantitative assessment of higher loop orders 
is required to judge their importance. 
In preparing such an assessment, we review bubble diagrams up to 
two-loop order in Sec.\,\ref{SEC:3}. In a next step, 
Sec.\,\ref{SEC:4} and App.\,\ref{app1} address the computational 
intricacies one encounters in computing the 2PI 
three-loop diagram. Namely, we derive the expression for the (six-dimensional) 
loop integral and investigate the vertex constraint to be able to set up a high-temperature 
treatment of this integral. While low temperatures can be subjected to 
the Monte-Carlo method, we are able to analytically extract the leading powers in the 
high-temperature situation for each possible scattering-channel 
combination (diagonal and off-diagonal): The diagonal case exhibits a leading-power thirteen temperature 
dependence while the off diagonal case contributes 
with maximal power four only. Compared to the high-temperature 
one-loop pressure there thus is a relative power of nine for the diagonal case 
which would ruin the idea of the thermal ground state and its 
free quasiparticle excitations to represent a useful 
a priori estimate if the loop expansion was to be truncated at 
order three. Fortunately, we had already shown in Sec.\,\ref{SEC:2} that this is not the 
entire story since there are infinitely many dihedrally symmetric 
diagrams which enjoy a finite support in their loop integrations. 
In Sec.\,\ref{SEC:5} we formally resum their contributions in terms of the according, truncated 
Dyson-Schwinger equation for the full 4-vertex. This yields an analytical continuation of the 
well-controlled low-temperature situation which turns out to be extremely well bounded. 
If the high-temperature form factor for the 4-vertex, as it is obtained from the Dyson-Schwinger treatment, 
is used to compute the leading temperature powers in two-loop and 2PI three-loop vertex 
resummed diagrams, then we are able to demonstrate that their leading (highly negative) 
powers cancel exactly. Also, for this form factor 
a smooth interpolation from extreme high-temperature suppression 
to unity at low temperatures suggests itself from the 
high-temperature behaviour already. In Sec.\,\ref{SEC:6} we summarise 
our results and provide some outlook on future work.

\section{\label{SEC:2}Exclusion of sign configurations in massive bubble diagrams}

In this section we count the number of energy-sign and scattering-channel combinations allowed by the constraints on momentum transfer in 4-vertices within 2PI diagrams that represent certain radiative corrections to the one-loop pressure, $P|_\mathrm{1-loop}$, up 
to six loops. For the according Feynman rules in the massive sector, including these constraints, see \cite{Hofmann2016}. 
By 2PI we understand those bubble diagrams that become 1-particle irreducible (1PI) 
contributions to the polarisation tensor upon cutting any single line. 

A useful observation \cite{hofmann-krasowski} concerning the momentum transfer constraints through 4-vertices 
is that the on-shellness of massive modes, $p^2=m^2=4e^2|\phi|^2$ ($|\phi|$ being the modulus of the inert, adjoint scalar 
field associated with densely packed (anti)caloron centers in the thermal ground state 
\cite{Hofmann2016,Entropy2016} setting the scale of maximal resolution) fixes 
the relative signs of the zero-components of four-momenta involved in the 
scattering process. For instance
\begin{align}\label{eq:generalconstraints1}
|(p+q)^2|\leq|\phi|^2 \quad &\Rightarrow \quad \sgn(p_0)=-\sgn(q_0)\,,\\
\label{eq:generalconstraints2}
|(p-q)^2|\leq|\phi|^2 \quad &\Rightarrow \quad \sgn(p_0)=\sgn(q_0)\,.
\end{align}
We here would like to investigate how these constraints decrease 
the number of sign configurations for loop momenta as the number of loops is 
increased in 2PI bubble diagrams contributing to the pressure. Subsequently, this 
general scheme is applied to 2PI bubble diagrams up 
to six loops: The ratio $R$ of the allowed number of energy-sign and scattering-channel combinations 
to such a priori possible combination is computed for each diagram. 
We find that $R$ 
decreases monotonically with increasing loop order but never vanishes. Indeed, we 
shall prove that within the class of dihedrally symmetric 2PI bubble diagrams $R>0$ at any finite 
loop order. The possibility to demonstrate termination of the loop expansion at a 
finite irreducible order \cite{hofmann-krasowski} is therefore ruled out. Loosely speaking, 
this relates to the fact that the more symmetric bubble diagrams are the less independent their 
vertex constraints become. This very fact is ignored in the naive counting of the number of 
constraints per loop variable as performed in \cite{hofmann06}.     

\subsection{Exclusion tables}\label{sec:systematics}
 \ytableausetup{aligntableaux=bottom}

To analyse excluded sign configurations in a systematic way the following combinatorics applies: 
Let $n$ be the number of vertices and $N=2n$ the number of internal lines. 
There are $3^n$ a priori possible combinations of scattering channels (distinguished by invariant momentum transfer in the Mandelstam variables $s$,$t$, and $u$) subject to their 
corresponding constraints. Moreover, there are $2^N$ possible sign combinations  
for the zero-components in loop four-momenta. (The modulus of such a zero-component is bounded by 
mass $m$.) In analogy to \cite{hofmann-krasowski} we organise all possible sign combinations 
in a table -- one table for each vertex and scattering channel -- and mark the excluded ones by gray elements. In this way, one obtains $3n$ tables. The following table format is convenient. 
For $N$ loop momenta, all zero-components can be positive ($\binom{N}{0}=1$ configurations, first row), only one 
zero-component is negative ($\binom{N}{1}=N$ configurations, second row), two zero-components are negative 
($\binom N2$ configurations, third row), etc. Thus, there are $N+1$ rows in the table, the 
$k$-th row possessing $\binom N{k-1}$ columns. 
For instance, setting $n=2$ there are four momenta $a$, $b$, $c$, and $d$. Thus, we assign the following a priori table:\\ 
\eqb
\label{table-gen}
\begin{ytableau}
\emptyset\\
a&b&c&d\\
ab&ac&ad&bc&bd&cd\\
a&b&c&d\\
\emptyset\\
\end{ytableau}\,.
\eqe
In the first row the symbol $\emptyset$ denotes that $a_0>0$ and $b_0>0$ and $c_0>0$, and $d_0>0$, in the second row either 
$a_0<0$ or $b_0<0$ or $c_0<0$, or $d_0<0$ and the others are positive, etc. Note that 
the constraints \eqref{eq:generalconstraints1} and \eqref{eq:generalconstraints2} determine relative signs only. Therefore, 
the exclusion tables are symmetric about the central row, and it is sufficient to consider the first $N/2+1$ rows only. 
Moreover, the $(N/2+1)$th row is symmetric about its center: 
The first entry is equivalent to the last, the second entry is equivalent to the second-to-last, and so forth. Therefore, the number of elements in the $(N/2+1)$th row can be halved. The overall-sign redundancy thus is completely removed 
by halving the number of elements in the table.

As an example, let us consider the $s$-channel constraint on the four-momenta $a$, $b$, $c$, and $d$
\[
|(a+b)^2|=|(c+d)^2|\leq|\phi|^2\ \Rightarrow\ \sgn{(a_0)}=-\sgn{(b_0)}\,,\ \sgn{(c_0)}=-\sgn{(d_0)}\,.
\]
At least two signs must differ from one another, ruling out row 1 and 2 (gray elements). In row 3 all combinations in which $\sgn{(a_0)}=\sgn{(b_0)}$ or $\sgn{(c_0)}=\sgn{(d_0)}$ are excluded. That is, $a_0>0$ and $b_0>0$ does not occur. 
This leaves only two options in total:\\
\eqb
\label{table-4}
\begin{ytableau}
*(gray)\\
*(gray)&*(gray)&*(gray)&*(gray)\\
*(gray)&ac&ad\\
\end{ytableau}\,.
\eqe
One proceeds by overlaying the tables for different vertices subject to given scattering channels. 
For a given element in the final table the 
overlay process enforces gray if at least one of the according elements in the contributing tables is gray. For example, a 
diagram with $n=3$ would or would not comprise the excluded combination $s$-channel on vertex 1, $u$-channel on 
vertex 2, and $u$-channel on vertex 3 ($suu$) if an overlay of the exclusion 
tables for each associated vertex produced an entirely or partially gray table, respectively.

\subsection{Exclusions for 2PI pressure diagrams up to six loops}

For the radiative corrections due to the massive sector of deconfining SU(2) Yang-Mills thermodynamics bubble diagrams and their symmetry factors are identical to those of $\lambda\varphi^4$-theory \cite{kleinert}. Below we treat all 2PI diagrams up to six-loop order, and, due to limited space, we state explicit exclusion tables up to four loops only. 
For the computation of excluded sign combinations in five- and 
six-loop diagrams we resort to a Mathematica code whose notebook is available upon request 
from the authors. This code is also listed in \cite{BischerMA2017}. It yields the ratio $R$ of allowed 
vs. $3^n\cdot 2^{2n-1}$ a priori possible (non-redundant) energy-sign and scattering-channel 
configurations\footnote{There are $3^n$ channel combinations, each carrying $2^{N-1}=2^{2n-1}$ energy-sign combinations.}.

\subsubsection{Two-loop diagram}\label{sec:2-loop-sign}
\begin{figure}
\centering{
{\Large $\frac{1}{8}\cdot\,$}
\includegraphics[align=c]{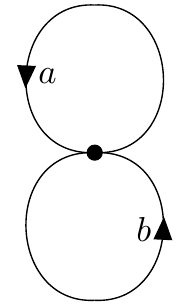}
\phantom{{\Large $\frac{1}{8}\cdot\,$}}
\caption{The only two-loop diagram (symmetry factor $1/8$).}
\label{fig:2-loop}
}
\end{figure}
The two-loop diagram in \autoref{fig:2-loop} is subject to one vertex constraint only and was calculated in 
\cite{herbst-hofmann-rohrer,SGH2007}. There is only one allowed sign-configuration: $\sgn(a_0)=-\sgn(b_0)$ in this case. The modulus of the contribution from \autoref{fig:2-loop} divided by the 1-loop result in the massive sector is less than $3.5\times$\num{5E-6} everywhere in the deconfining phase.

\subsubsection{2PI three-loop diagram}\label{sec:3-loop-signs}
\begin{figure}[H]
\centering{
{\Large $\frac{1}{48}\cdot\,$}
\includegraphics[align=c]{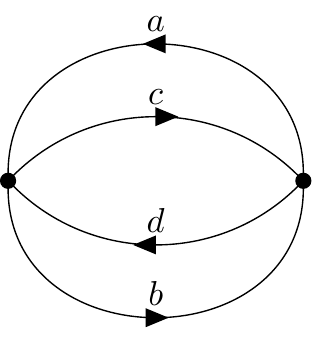}
\phantom{{\Large $\frac{1}{48}\cdot\,$}}
\caption{The only 2PI three-loop diagram (symmetry factor $1/48$).}
\label{fig:3-loop-1}
}
\end{figure}
For the three-loop diagram in \autoref{fig:3-loop-1} the a priori table is\\ 
\eqb
\label{table-3loop}
\begin{ytableau}
\emptyset\\
a&b&c&d\\
ab&ac&ad\
\end{ytableau}\,,
\eqe
and both vertices are subject to the same set of $s$-, $t$-, and $u$-constraints. Namely, 
\begin{align}
s=|(a+d)^2|=|(b+c)^2|\leq|\phi|^2\ &\Rightarrow \sgn{(a_0)}=-\sgn{(d_0)},\, \sgn{(b_0)}=-\sgn{(c_0)} \label{eq:3-constraint-s}\\ 
t=|(a-b)^2|=|(c-d)^2|\leq|\phi|^2\ &\Rightarrow \sgn{(a_0)}=\sgn{(b_0)},\, \sgn{(c_0)}=\sgn{(d_0)} \label{eq:3-constraint-t}\\
u=|(a-c)^2|=|(d-b)^2|\leq|\phi|^2\ &\Rightarrow \sgn{(a_0)}=\sgn{(c_0)},\, \sgn{(d_0)}=\sgn{(b_0)}\label{eq:3-constraint-u}\,.
\end{align}
This implies the following exclusion tables \ytableausetup{smalltableaux} \\\\
s: \begin{ytableau}
*(gray)\\
*(gray)&*(gray)&*(gray)&*(gray)\\
&&*(gray)\\
\end{ytableau}
t: \begin{ytableau}
\\
*(gray)&*(gray)&*(gray)&*(gray)\\
&*(gray)&*(gray)\\
\end{ytableau}
u: \begin{ytableau}
\\
*(gray)&*(gray)&*(gray)&*(gray)\\
*(gray)&&*(gray)\\
\end{ytableau}\,.\\\\
By overlaying two such tables one infers that for $ss$, $tt$, and $uu$ there are two allowed 
sign configurations,  
\eqb
\label{3-loopallowedsigns-diag}
n(ss)=n(tt)=n(uu)=2\,.
\eqe
For $st$, $su$, and $tu$ only one allowed sign configuration 
remains, 
\eqb
\label{3-loopallowedsigns-nondiag}
n(st)=n(ts)=n(su)=n(us)=n(tu)=n(ut)=1\,.
\eqe
The numerator in the ratio $R$ thus computes as 
\[
\begin{split}
n(ss)+n(tt)+n(uu)+n(st)+n(ts)+n(su)+n(us)
+n(tu)+n(ut) = 12\,.
\end{split}
\]
Therefore, one obtains 
\eqb
\label{R3loop}
R=12/(9\cdot8)=1/6
\eqe
for the diagram in \autoref{fig:3-loop-1}. 

\subsubsection{2PI four-loop diagram}
\begin{figure}[H]
\centering{
{\Large $\frac{1}{48}\cdot$}
\includegraphics[align=c]{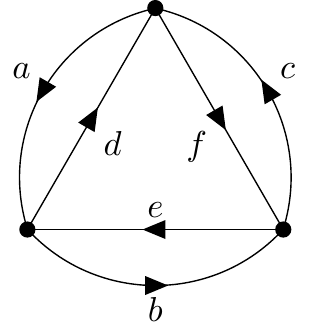}
\phantom{{\Large $\frac{1}{48}\cdot$}}
\caption{The only 2PI four-loop diagram (symmetry factor $1/48$).}
\label{fig:4-loop-1}
}
\end{figure}
For the four-loop diagram in \autoref{fig:4-loop-1} we use the following 
channel conventions at each of the vertices $V_1$, $V_2$, and $V_3$\\ 
$V_1$:
\begin{align}
s=|(a+e)^2|=|(b+d)^2|\leq|\phi|^2\ &\Rightarrow \sgn{(a_0)}=-\sgn{(b_0)},\,\sgn{(c_0)}=-\sgn{(d_0)} \\
t=|(a-b)^2|=|(e-d)^2|\leq|\phi|^2\ &\Rightarrow \sgn{(a_0)}=\sgn{(c_0)},\,\sgn{(b_0)}=\sgn{(d_0)} \\
u=|(a-d)^2|=|(e-b)^2|\leq|\phi|^2\ &\Rightarrow \sgn{(a_0)}=\sgn{(d_0)},\,\sgn{(b_0)}=\sgn{(c_0)}\,,
\end{align}
$V_2$:
\begin{align}
s=|(b+f)^2|=|(c+e)^2|\leq|\phi|^2 \ &\Rightarrow \sgn{(e_0)}=-\sgn{(b_0)},\, \sgn{(f_0)}=-\sgn{(d_0)} \\
t=|(b-c)^2|=|(f-e)^2|\leq|\phi|^2 \ &\Rightarrow \sgn{(f_0)}=\sgn{(e_0)},\, \sgn{(b_0)}=\sgn{(d_0)} \\
u=|(b-e)^2|=|(f-c)^2|\leq|\phi|^2 \ &\Rightarrow \sgn{(e_0)}=\sgn{(d_0)},\, \sgn{(b_0)}=\sgn{(f_0)}\,,
\end{align}
$V_3$:
\begin{align}
s=|(c+d)^2|=|(a+f)^2|\leq|\phi|^2 \ &\Rightarrow \sgn{(c_0)}=-\sgn{(e_0)},\, \sgn{(f_0)}=-\sgn{(a_0)} \\
t=|(c-a)^2|=|(d-f)^2|\leq|\phi|^2 \ &\Rightarrow \sgn{(a_0)}=\sgn{(c_0)},\, \sgn{(f_0)}=\sgn{(e_0)} \\
u=|(c-f)^2|=|(d-a)^2|\leq|\phi|^2 \ &\Rightarrow \sgn{(c_0)}=\sgn{(f_0)},\, \sgn{(e_0)}=\sgn{(a_0)}\,.
\end{align}
The a priori table thus reads \ytableausetup{nosmalltableaux}\\
\eqb
\label{ApT4}
\begin{ytableau}
\emptyset\\
a&b&c&d&e&f\\
ab&ac&ad&ae&af&bc&bd&be&bf&cd&ce&cf&de&df&ef\\
abc&abd&abe&abf&acd&ace&acf&ade&adf&aef\\
\end{ytableau}\,.
\eqe\\ 
Based on this table we obtain the following exclusion tables for each vertex and scattering channel 
\ytableausetup{smalltableaux}\\\newpage

$V_1s$: \begin{ytableau}
*(gray)\\
*(gray)&*(gray)&*(gray)&*(gray)&*(gray)&*(gray)\\
&*(gray)&&*(gray)&*(gray)&*(gray)&*(gray)&&*(gray)&*(gray)&*(gray)&*(gray)&&*(gray)&*(gray)\\
&*(gray)&*(gray)&&&*(gray)&*(gray)&*(gray)&&*(gray)\\
\end{ytableau}\\\\

$V_1t$: \begin{ytableau}
\\
*(gray)&*(gray)&&*(gray)&*(gray)&\\
&*(gray)&*(gray)&*(gray)&*(gray)&*(gray)&*(gray)&*(gray)&*(gray)&*(gray)&*(gray)&&&*(gray)&*(gray)\\
&*(gray)&*(gray)&&*(gray)&*(gray)&*(gray)&*(gray)&*(gray)&*(gray)\\
\end{ytableau}\\\\

$V_1u$: \begin{ytableau}
\\
*(gray)&*(gray)&&*(gray)&*(gray)&\\
*(gray)&*(gray)&&*(gray)&*(gray)&*(gray)&*(gray)&&*(gray)&*(gray)&*(gray)&&*(gray)&*(gray)&*(gray)\\
*(gray)&*(gray)&*(gray)&*(gray)&&*(gray)&*(gray)&*(gray)&&*(gray)\\
\end{ytableau}\\\\

$V_2s$: \begin{ytableau}
*(gray)\\
*(gray)&*(gray)&*(gray)&*(gray)&*(gray)&*(gray)\\
*(gray)&*(gray)&*(gray)&*(gray)&*(gray)&&*(gray)&&*(gray)&*(gray)&*(gray)&&*(gray)&*(gray)&\\
&*(gray)&&*(gray)&*(gray)&*(gray)&&*(gray)&*(gray)&\\
\end{ytableau}\\\\

$V_2t$: \begin{ytableau}
\\
&*(gray)&*(gray)&&*(gray)&*(gray)\\
*(gray)&*(gray)&&*(gray)&*(gray)&&*(gray)&*(gray)&*(gray)&*(gray)&*(gray)&*(gray)&*(gray)&*(gray)&\\
&*(gray)&*(gray)&*(gray)&*(gray)&*(gray)&*(gray)&*(gray)&*(gray)&\\
\end{ytableau}\\\\

$V_2u$: \begin{ytableau}
\\
&*(gray)&*(gray)&&*(gray)&*(gray)\\
*(gray)&*(gray)&&*(gray)&*(gray)&*(gray)&*(gray)&&*(gray)&*(gray)&*(gray)&&*(gray)&*(gray)&*(gray)\\
*(gray)&*(gray)&&*(gray)&*(gray)&*(gray)&&*(gray)&*(gray)&*(gray)\\
\end{ytableau}\\\\

$V_3s$: \begin{ytableau}
*(gray)\\
*(gray)&*(gray)&*(gray)&*(gray)&*(gray)&*(gray)\\
*(gray)&&&*(gray)&*(gray)&*(gray)&*(gray)&*(gray)&*(gray)&*(gray)&*(gray)&&*(gray)&&*(gray)\\
&&*(gray)&*(gray)&*(gray)&&*(gray)&&*(gray)&*(gray)\\
\end{ytableau}\\\\

$V_3t$: \begin{ytableau}
\\
*(gray)&&*(gray)&*(gray)&&*(gray)\\
*(gray)&&*(gray)&*(gray)&*(gray)&*(gray)&*(gray)&&*(gray)&*(gray)&*(gray)&*(gray)&*(gray)&&*(gray)\\
&*(gray)&*(gray)&*(gray)&*(gray)&&*(gray)&*(gray)&*(gray)&*(gray)\\
\end{ytableau}\\\\

$V_3u$: \begin{ytableau}
\\
*(gray)&&*(gray)&*(gray)&&*(gray)\\
*(gray)&*(gray)&&*(gray)&*(gray)&*(gray)&*(gray)&&*(gray)&*(gray)&*(gray)&&*(gray)&*(gray)&*(gray)\\
*(gray)&&*(gray)&*(gray)&*(gray)&*(gray)&*(gray)&&*(gray)&*(gray)\ \ \ \ \ \ \ \ \ \ \ \ \ \ \ \ \,.\\ 
\end{ytableau}\\\\\\
The allowed sign configurations for all 27 possible channel combinations are obtained by overlaying the respective tables and identifying the entries which are white thereafter. For example, the allowed sign configurations of the $sss$-channel are obtained by overlaying the tables $V_1s$, $V_2s$ and $V_3s$, where one finds that only the entry in the bottom left corner (with label $abc$ in \eqref{ApT4}) is white in all three tables. In \autoref{tab:4-loop-1} all channel combinations and their 
allowed energy-sign combinations, denoted by their label in the a priori table (\ref{ApT4}), are listed.
\begin{table}
\begin{tabu}{X[$]X[$]|X[$]X[$]|X[$]X[$]}
\toprule
$Channels$ & $Allowed$ &$Channels$ & $Allowed$ &$Channels$ & $Allowed$\\
\midrule
sss&abc &tss&cf,\, abc&uss&cf   \\
sst&be,\,abc &tst&abc&ust&be   \\
ssu&be &tsu&cf&usu&be,\,cf  \\
sts&ad,\,abc  &tts&abc &uts&ad   \\
stt&abc &ttt&\emptyset,\,abc&utt&\emptyset\\
stu&ad &ttu&\emptyset& utu& \emptyset,\,ad  \\
sus&ad& tus & cf& uus &ad,\,cf\\
sut&be &tut & \emptyset & uut & \emptyset,\,be   \\
suu&ad,\,be & tuu & \emptyset,\,cf & uuu & \emptyset, \, ad,\,be,\,cf    \\
\bottomrule
\end{tabu}
\caption{Scattering channels and their respectively allowed sign configurations for the 2PI four-loop diagram in \autoref{fig:4-loop-1}. Sign configurations are denoted by their label in the 2PI four-loop a priori table \eqref{ApT4}. By $\mathbb{Z}_2$-symmetry (over-all sign flip), each label corresponds to \emph{two} actual sign configurations. This means that, for instance, $\emptyset$ denotes the configurations with all six zero components having the same sign, $be$ denotes the configurations ${b^0,e^0>0>a^0,c^0,d^0,f^0}$ and ${b^0,e^0<0<a^0,c^0,d^0,f^0}$, while $abc$ denotes the configurations ${a^0,b^0,c^0>0>d^0,e^0,f^0}$ and ${a^0,b^0,c^0<0<d^0,e^0,f^0}$ and so forth (cf. Sec.\,\ref{sec:systematics}) .}
\label{tab:4-loop-1}
\end{table}
Using this table, we conclude that
\eqb
\label{R4loop}
R=\frac{40}{864}=\frac{5}{108}=0.0463\,.
\eqe

\subsubsection{The two 2PI five-loop diagrams}\label{5loop}

\begin{figure}[H]
{\centering{{\Large$\frac{1}{128}\cdot$}
\includegraphics[scale=1,align=c]{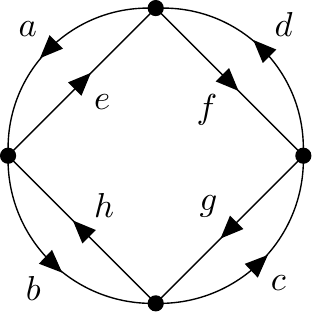}
\phantom{{\Large$\frac{1}{128}\cdot$}}
\qquad
{\Large$\frac{1}{32}\cdot$}
\includegraphics[scale=1,align=c]{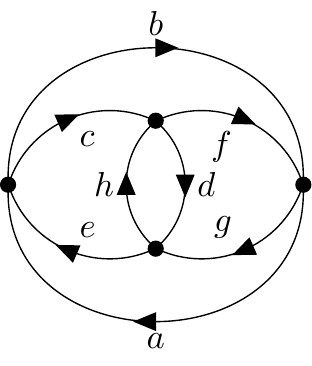}
\phantom{{\Large$\frac{1}{128}\cdot$}}
\caption{The first and second 2PI five-loop diagram (symmetry factors $1/128$ (first) and $1/32$ (second)).}
\label{fig:5-loop}}}
\end{figure}
The two 2PI five-loop diagrams are shown in \autoref{fig:5-loop}. Instead of explicitly stating and comparing tables, which becomes  increasingly tedious, we refer from here on to a Mathematica code to compute the ratio $R$. 
For the first and most symmetric diagram in \autoref{fig:5-loop} one obtains
\eqb
\label{R5loopl}
R=1/72=0.0139\,.
\eqe
For the second diagram in \autoref{fig:5-loop} one computes
\eqb
\label{R5loopr}
R=1/81=0.0123\,.
\eqe

\subsubsection{The four 2PI six-loop diagrams}\label{6loop}

\begin{figure}[H]
{\centering{{\Large$\frac{1}{320}\cdot$}
\includegraphics[scale=1,align=c]{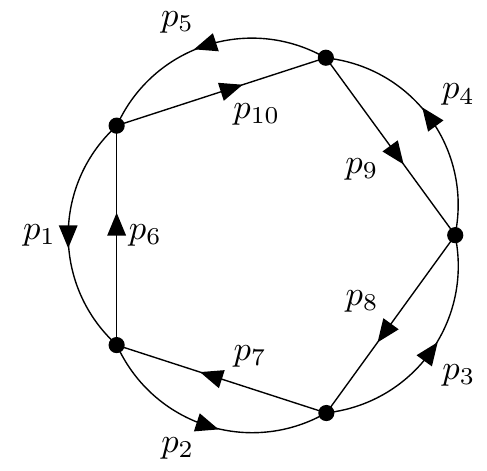}
\phantom{{\Large$\frac{1}{320}\cdot$}}
\caption{The first and most symmetric 2PI six-loop diagram (symmetry factor $1/320$).}
\label{fig:6-loop-1}}}
\end{figure}

The four 2PI six-loop diagrams are shown in \autoref{fig:6-loop-1} and \autoref{fig:6-loop-2-4}.
For the most symmetric diagram in \autoref{fig:6-loop-1} one obtains 
\eqb
\label{R6loop1}
R=17/3888=0.0044\,,
\eqe
for the other diagrams shown in \autoref{fig:6-loop-2-4}, one has in the order of appearance 
\eqb
\label{R6loop2}
R=7/1944=0.0036\,,
\eqe
\eqb
\label{R6loop3}
R=13/3888=0.0033\,,
\eqe
and 
\eqb
\label{R6loop4}
R=1/324=0.0031\,.
\eqe

\subsubsection{Summary of results up to six loops}

\autoref{tab:signconstraints} summarises our results for $R$ and, as an aside, for the product $R\cdot S$, $S$ denoting 
the symmetry factor of a given diagram, up to 2PI six-loop order. Notice that while $R$ drops with increasing loop 
order it is always finite. In fact, we prove in Sec.\,\ref{sec:n-gon-theorem} that there exists a class of 2PI 
diagrams (those with dihedral symmetry) for which $R>0$ at any finite loop order.   
\begin{subfigures}
\begin{figure}
{\centering{
{\Large$\frac{1}{32}\cdot$}
\includegraphics[align=c]{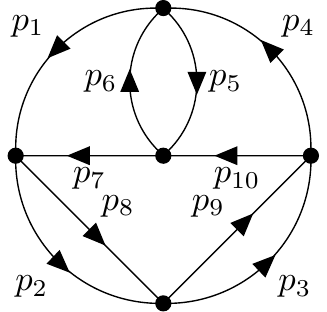}
\phantom{{\Large$\frac{1}{32}\cdot$}}
\qquad
{\Large$\frac{1}{16}\cdot$}
\includegraphics[align=c]{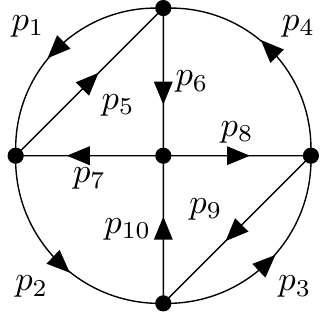}
\phantom{{\Large$\frac{1}{16}\cdot$}}
\caption{The second and third 2PI six-loop diagrams (symmetry factors $1/32$ and $1/16$).}
\label{fig:6-loop-2}}}
\end{figure}
\begin{figure}
{\centering{
{\Large$\frac{1}{120}\cdot$}
\includegraphics[align=c]{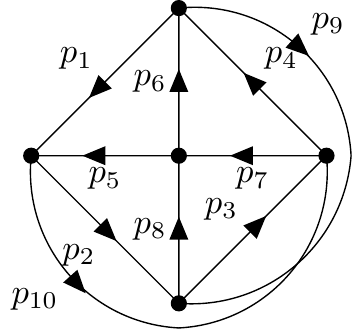}
\phantom{{\Large$\frac{1}{120}\cdot$}}
\caption{The fourth 2PI six-loop diagram (symmetry factor $1/120$). This is the only non-planar diagram up 
to six-loop order.}
\label{fig:6-loop-3}}}
\end{figure}
\label{fig:6-loop-2-4}
\end{subfigures}

\begin{table}
\begin{tabu}{rrX[r]X[r,$]X[r, $]}
\toprule
Loop number & Diagram number & $R$ & S^{-1} & R\cdot S\\
\midrule
3	&1	&0.1667 & 48 & 0.00347222\\\addlinespace[2pt]
4	&1	&0.0463 & 48 & 0.00096451\\\addlinespace[2pt]
5	&1	&0.0139 & 128 & 0.00010851\\
5	&2	&0.0123 & 32 & 0.00038580\\\addlinespace[2pt]
6	&1	&0.0044 & 320 & 0.00001366\\
6	&2	&0.0036 & 32 & 0.00011253\\
6	&3	&0.0033 & 16 & 0.00020898\\
6	&4	&0.0033 & 120 & 0.00002572\\
\bottomrule
\end{tabu}
\caption{Ratio $R$ of allowed vs. a priori possible energy-sign and scattering-channel combinations 
for 2PI bubble diagrams up to six loops. $S$ denotes a diagram's symmetry factor.}
\label{tab:signconstraints}
\end{table}

\subsection{Finiteness of $R$ at any finite loop order}\label{sec:n-gon-theorem}

Here we give a proof by induction showing that $R>0$ at any finite loop order within
the class $\mathcal{C}$ of diagrams symmetric under the dihedral group. That is, we address diagrams 
with the corners of an $n$-gon touching the enclosing circle. In other words, 
these diagrams represent a closed chain of pairs of momenta, tied together by $n$ vertices. 
Examples are \autoref{fig:4-loop-1} and the left diagram in \autoref{fig:5-loop}.
Even though there is no such thing as a $2$-gon, we admit the three-loop diagram in \autoref{fig:3-loop-1} 
to this class.

We now prove that any diagram in class $\mathcal{C}$ is subject to \emph{one independent constraint only} provided that its vertices exclusively convey $u$-channel scattering. To set up the induction step, consider the case $n\geq3$. We make the convention that momenta along the circle flow counter-clockwise and that momenta along the $n$-gon flow clockwise: 
\begin{center}
\includegraphics{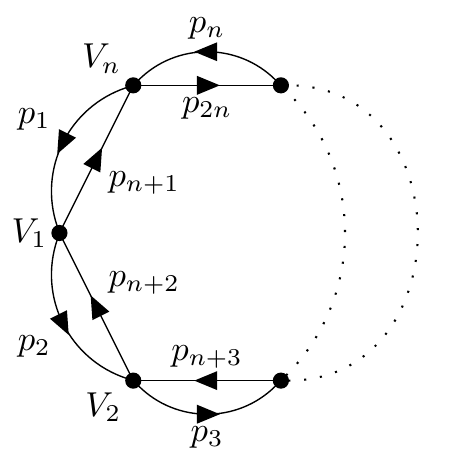}
\end{center}
We define the $u$-channel momentum within a given vertex by subtracting the outgoing $n$-gon momentum from the ingoing circle momentum. According to this convention, the $k$-th vertex is $u$-channel constrained as
\[\label{eq:induction-a}
|(p_k-p_{n+k})^2|=|(p_{n+k+1}-p_{k+1})^2|\leq |\phi|^2\,,
\]
where $k=1,\dots,n-1$ and
\[\label{eq:induction-b}
|(p_n-p_{2n})^2|=|(p_{n+1}-p_{1})^2|\leq|\phi|^2
\]
for $k=n$. These constraints are redundant by momentum conservation at each vertex. Namely,
\[
\begin{split}\label{eq:n-gon-chain}
|(p_1-p_{n+1})^2|&=|(p_{n+2}-p_{2})^2|=|(p_{2}-p_{n+2})^2|=|(p_{n+3}-p_{3})^2|=\dots =\\
&= |(p_n-p_{2n})^2|=|(p_{n+1}-p_1)^2|=|(p_1-p_{n+1})^2|\leq|\phi|^2\,.
\end{split}
\]
To perform the induction step, we pinch in an extra vertex $V_{n+1}$ inbetween $V_n$ and $V_1$.
This invokes new momenta $q$ and $s$ while $p_1$ and $p_{n+1}$ now link $V_1$ and $V_{n+1}$.
\begin{center}
\includegraphics{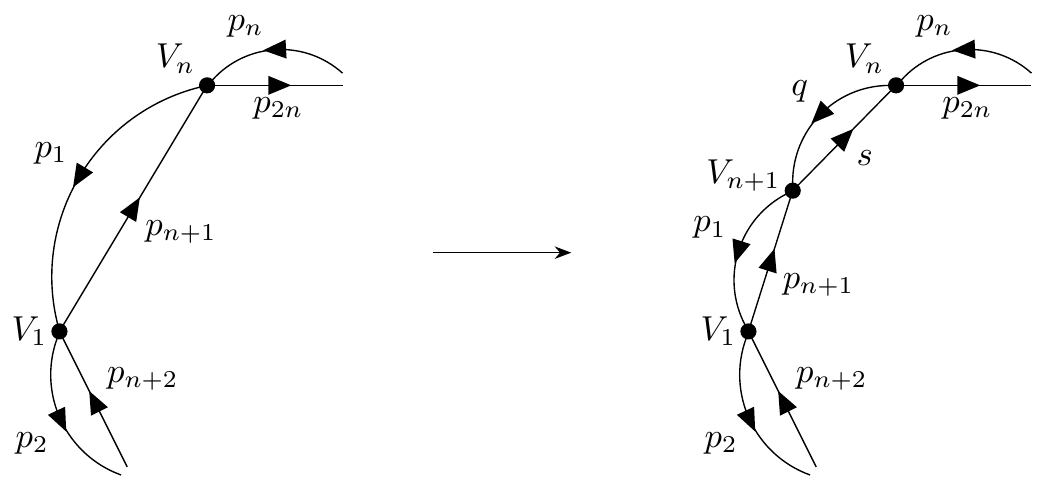}
\end{center}
Notice that the constraints on $V_1$ up to $V_{n-1}$ remain unchanged. 
On the other hand, the constraints on $V_n$ and $V_{n+1}$ read
\begin{align}
V_n:\quad &|(p_n-p_{2n})^2|=|(s-q)^2|\leq|\phi|^2\\
V_{n+1}:\quad &|(q-s)^2|=|(p_{n+1}-p_1)^2|\leq|\phi|^2\,.
\end{align}
Thus, the chain in \eqref{eq:n-gon-chain} can be continued up to $V_{n+1}$,
\[
\begin{split}\label{eq:n-gon-chain2}
|(p_1-p_{n+1})^2|&=|(p_{n+2}-p_{2})^2|=\dots =\\
&= |(p_n-p_{2n})^2|=|(s-q)^2|=|(p_{n+1}-p_{1})^2|\leq|\phi|^2\,.
\end{split}
\]
Again, the extended chain is subject to \emph{one independent constraint only}. It is obvious that the case $n=3$ (cf. \autoref{fig:4-loop-1}) shown below represents a valid induction base.
\begin{figure}[H]
\begin{center}\includegraphics{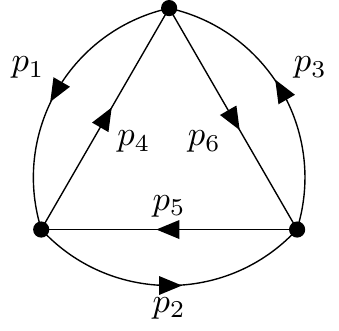}\end{center}
\end{figure}
As a consequence, the support of the according loop integration is finite, independently of $n$. In turn, this implies $R>0$ for all diagrams within class $\mathcal{C}$ which refutes the conjecture voiced in \cite{hofmann06} that convergence of the 
2PI loop expansion occurs by termination at a finite loop order. We are thus motivated to perform a more quantitative 
analysis of the diagrams in class $\mathcal{C}$, beginning at three-loop order. As we will demonstrate, the 
three-loop case splits into a contribution, which is subject to a single vertex constraint only, as well as a piece being subject to two independent constraints. The power of vertex constraints is impressively demonstrated by the a difference in leading power in $\lambda$ of nine.   

\section{\label{SEC:3}Review of up to two-loop diagrams}

As an interlude and to prepare the investigations performed in Secs.\,\ref{SEC:4} 
and \ref{SEC:5} we here review results on one-loop and two-loop diagrams 
within the massive sector of deconfining SU(2) Yang-Mills thermodynamics 
\cite{SGH2007}.   

\subsection{One-loop pressure}

On the level of free thermal quasiparticle excitations, that is, on the one-loop level the pressure exerted 
by the massive sector reads\footnote{We only quote 
that part of the pressure 
which is represented by excitations, the thermal ground state contributes a 
negative portion which is not considered here.}  \cite{Hofmann2016}   
\begin{align}
\label{eq:oneloop-pressuremassive}
P(\lambda)|_\mathrm{1-loop} &= -\Lambda^4\frac{12\lambda^4}{(2\pi)^6}\bar P(2a)\,, 
\end{align}
where 
\begin{align}\label{eq:dimpress}
\bar P(y) &=\int_0^\infty \der x\, x^2\log\left[1 - e^{-\sqrt{x^2+y^2}}\right]\,,
\end{align} 
$\Lambda$ denotes the Yang-Mills scale, $\lambda\equiv \frac{2\pi T}{\Lambda},$ and $a\equiv\frac{m}{2T}$.

\subsection{Two-loop correction}
\begin{figure}
\centering{
{\Large $\frac{1}{8}\cdot\,$}
\includegraphics[align=c]{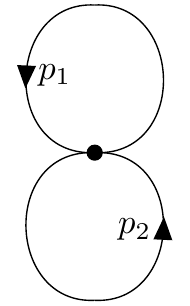}
\phantom{{\Large $\frac{1}{8}\cdot\,$}}
\caption{The two-loop diagram for the pressure in the massive sector of deconfining SU(2) Yang-Mills thermodynamics (symmetry factor $1/8$).}
\label{fig:2-loop-2}
}
\end{figure}
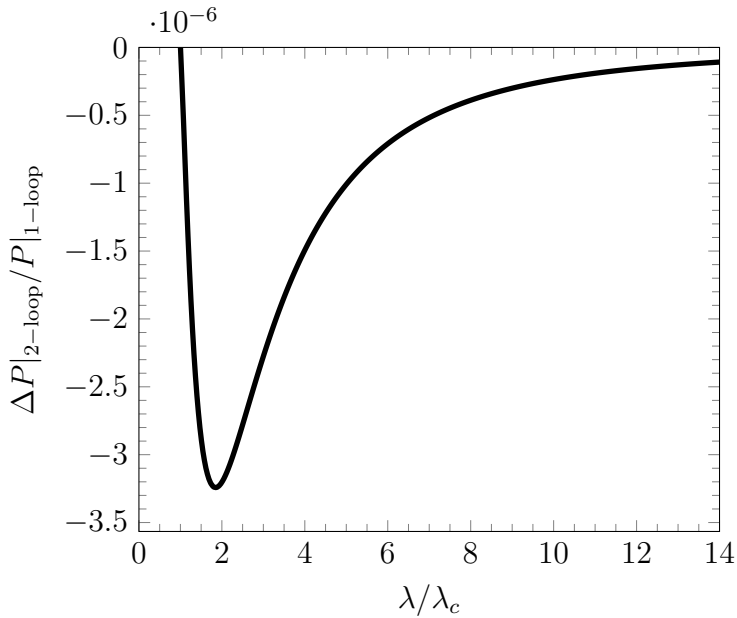
\begin{figure}
\centering{
    \begin{tikzpicture}
      \begin{axis}[
          width=0.6\linewidth, 
          xlabel=$\lambda/\lambda_c$, 
          ylabel=$\Delta P|_{\mathrm{2-loop}}/P|_{\mathrm{1-loop}}$,
          minor x tick num={3},
          minor y tick num={4},
          xmin=0, xmax=14,
         ymax=0,
        ]
        \addplot[
        smooth,
        line width=2pt,
        ] 
        table{graphs/2-loop-plot.dat}; 
      \end{axis}
    \end{tikzpicture}
    \hphantom{$-1.5$}
\caption{Plot of the ratio of two-loop vs. one-loop contributions to the pressure arising from the massive sector of deconfining SU(2) Yang-Mills thermodynamics as a function of dimensionless temperature $\lambda=2\pi \frac{T}{\Lambda}$. The deconfining phase extends down to $\lambda_c=13.87$.
}
\label{fig:2-loop-plot}
}
\end{figure}

The pressure contribution associated with the two-loop diagram in \autoref{fig:2-loop-2} reads \cite{herbst-hofmann-rohrer,SGH2007}
\begin{multline}\label{eq:2-loop-exact}
\Delta P|_\mathrm{2-loop} = \frac{-2e^2T^4}{\lambda^6}\intkern\der r_1\der r_2\der\cos\theta\frac{r_1^2r_2^2}{\sqrt{r_1^2+m^2}\sqrt{r_2^2+m^2}}\\
\times\left[ 14-2\frac{k^4}{m^4}\right]
 n_B\left(2\pi\sqrt{\frac{r_1^2+m^2}{\lambda^3}}\right)n_B\left(2\pi\sqrt{\frac{r_2^2+m^2}{\lambda^3}}\right) \,,
\end{multline}
where 
\[
k^2\equiv p_1p_2 = -\sqrt{r_1^2+m^2}\sqrt{r_2^2+m^2} -r_1r_2\cos\theta
\] 
is defined as the Lorentz-invariant product of the dimensionless\footnote{We normalise physical four-momentum components $P^\mu$ by $|\phi|$ to arrive at dimensionless components $p^\mu$. Likewise, the physical mass is made dimensionless: $m=2e$.}
loop four-momenta $p_1$ and $p_2$, $r_1=|\mathbf{p}_1|$ and $r_2=|\mathbf{p}_2|$ denote the moduli of their spatial parts, $n_B(x) = (\exp(x)-1)^{-1}$ refers to the Bose-Einstein distribution function, and 
the integration is subject to the constraint
\[
|2m^2-2\sqrt{r_1^2+m^2}\sqrt{r_2^2+m^2}-2r_1r_2\cos\theta|\leq1\,.
\]
In \autoref{fig:2-loop-plot} the temperature dependence of the numerical integrations in \eqref{eq:2-loop-exact} and \eqref{eq:oneloop-pressuremassive} is shown in terms of their ratio.

\section{\label{SEC:4}The 2PI three-loop diagram}

In the last section we have reviewed that the two-loop contribution to the pressure in the massive sector is strongly suppressed compared to the one-loop contribution for all temperatures. Because the results in \cite{KavianiHofmann} were based on an erroneous implementation of vertex constraints, a re-evaluation of the three-loop diagram in \autoref{fig:3-loop-1-b} is required. Interestingly, we find that a hierarchical suppression of three-loop vs. one-loop occurs only within a small band of low temperatures. As a consequence, the proper treatment of the radiative corrections induced by this particular diagram amounts to a resummation within an infinite class of higher loop diagrams. This resummation scheme will be addressed in Sec.\,\ref{SEC:5} in terms of a particular truncation of the Dyson-Schwinger (DS) hierarchy. The resulting resummed chain is well bounded and hierarchically smaller than the one-loop contributions for \emph{all} temperatures.

\subsection{\label{integral}Loop integral}
\begin{figure}
\centering{
{\Large $\frac{1}{48}\cdot\,$}
\includegraphics[align=c]{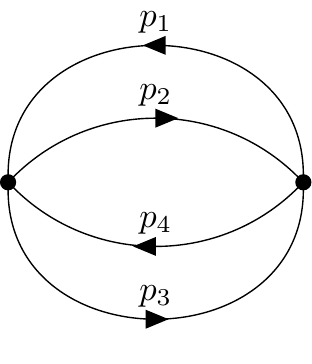}
\phantom{{\Large $\frac{1}{48}\cdot\,$}}
\caption{The 2PI three-loop diagram contributing to the pressure with symmetry factor $1/48$}
\label{fig:3-loop-1-b}
}
\end{figure}
Let us compute the diagram in \autoref{fig:3-loop-1-b}. To have an analytical grasp on it, we will perform a high-temperature expansion. Low temperatures are addressed by Monte Carlo methods. The consistency of both approaches is demonstrated by comparison within a transition regime.

To begin with, we quote the expression for the loop integral, derived in Appendix \ref{app1}, as
\begin{multline}\label{eq:3-loop-start}
\Delta P|_{\mathrm{3-loop}} = i\frac{\Lambda^4}{48\lambda^2}e^4\frac1{(2\pi)^6}\sum_{\mathrm{signs}}\intkern\der\theta_1\der\varphi_1\der r_1\der r_2\der\theta_3\sum_{\{r_3\}}r^2_1r^2_2r^2_3\sin\theta_1\sin\theta_3\\
\times P(p_i)\frac{n_B'(r_1)n_B'(r_2)n_B'(r_3)n_B'(r_4)}{8|p^0_1p^0_2p^0_3p^0_4|}\,.
\end{multline}
The first sum in \eqref{eq:3-loop-start} runs over allowed sign combinations for $p^0_i$, $i=1,\dots,4$, see Appendix \ref{app1}.
All four-momenta $p_i\equiv (p^0_i, \mathbf{p}_i)$ are on-shell, $|p_i^0|\equiv\sqrt{\mathbf{p}_i^2+m^2}$, and are parametrised as 
\begin{equation*}
p_4\equiv p_2+p_3-p_1,\quad \mathbf{p}_2\equiv\mat{0\\0\\r_2},\quad \mathbf{p}_3\equiv r_3\mat{0\\\sin\theta_3 \\\cos\theta_3}, \quad \mathbf{p}_1\equiv r_1\mat{\sin\theta_1\cos\varphi_1\\ \sin\theta_1\sin\varphi_1 \\ \cos\theta_1}\,.
\end{equation*}
In the equivalent cases $ss$, $tt$, $uu$ (diagonal), the integration is constrained by
\[\label{eq:constraint-ss}
|(p_1+p_4)^2|=|(p_2+p_3)|^2\leq 1\phantom{.}.
\]
Summing over these cases, the resulting contribution to $\Delta P|_{\mathrm{3-loop}}$ is denoted by $\linebreak[0]1/3\Delta P|_{\mathrm{3-loop},ss}$.
On the other hand, for the equivalent cases $st$, $su$, $tu$, $ts$, $us$, $ut$ (off-diagonal) the constraints on the integration read
\[\begin{split}\label{eq:constraint-st}
|(p_1+p_4)^2|&=|(p_2+p_3)|^2\leq 1\,,\\
|(p_1-p_2)^2|&=|(p_3-p_4)|^2\leq 1\,.
\end{split}\]
The sum of these cases amounts to $2/3\Delta P|_{\mathrm{3-loop},st}$ such that
\[\label{eq:pressure-3-loop}
\Delta P|_\mathrm{3-loop} = \frac13\Delta P|_{\mathrm{3-loop},ss} + \frac23\Delta P|_{\mathrm{3-loop},st}.
\]
The second sum in \eqref{eq:3-loop-start} runs over all solutions in $r_3$ of the equation
\begin{multline}\label{eq:r3}
\sgn(p^0_2)\sqrt{r_2^2+m^2}+\sgn(p^0_3)\sqrt{r^2_3+m^2}-\sqrt{r_1^2+m^2}=
-\left[r^2_1+r^2_2+r^2_3 \right.\\\left. - 2 r_1r_2\cos\theta_1  -2r_1r_3(\sin\varphi_1\sin\theta_1\sin\theta_3+\cos\theta_1\cos\theta_3)+2r_2r_3\cos\theta_3 + m^2\right]^{1/2}\,,
\end{multline}
see Appendix \ref{app1}.
The polynomial $P(\{p_i\})$reads
\begin{multline}\label{eq:polynomial}
P(\{p_i\})= 144 -12\frac1{m^4}\left\{\vphantom{\frac12}(p_1p_2)^2+(p_1p_3)^2+(p_1p_4)^2+(p_2p_3)^2+(p_2p_4)^2+(p_3p_4)^2 \right\} \\
+36\frac1{m^6}\left\{\vphantom{\frac12}(p_1p_2)(p_1p_3)(p_2p_3)+(p_1p_2)(p_1p_4)(p_2p_4)
\right.\\
\left.\vphantom{\frac12}
+(p_1p_3)(p_1p_4)(p_3p_4)+(p_2p_3)(p_2p_4)(p_3p_4)\right\}\\
+12\frac1{m^8}\left\{\vphantom{\frac12}(p_1p_2)^2(p_3p_4)^2+(p_1p_3)^2(p_2p_4)^2+(p_1p_4)^2(p_2p_3)^2 - (p_1p_2)(p_1p_3)(p_2p_4)(p_3p_4)
\right.\\
\left.\vphantom{\frac12} -(p_1p_2)(p_1p_4)(p_2p_3)(p_3p_4) - (p_1p_3)(p_1p_4)(p_2p_3)(p_2p_4)\right\},
\end{multline}
and the Bose-Einstein distribution shorthand notation is
\[
n_B'(r)\equiv n_B\left(2\pi\sqrt{r^2+m^2}/\lambda^{3/2}\right).
\]

In general, the constraints \eqref{eq:constraint-ss} and \eqref{eq:constraint-st} cannot explicitly be resolved. Thus in a numerical evaluation of the integral one has to sample the regions that satisfy conditions \eqref{eq:constraint-ss}, \eqref{eq:constraint-st} 
and that solve \eqref{eq:r3}, e.g. by the Monte Carlo method, while discarding complementary regions. However, as we demonstrate in Sec.\,\ref{sec:approximation}, the high-temperature limit lends itself to an analytical resolution of \eqref{eq:constraint-ss}, \eqref{eq:constraint-st},  and \eqref{eq:r3}.

\subsection{Properties of the vertex constraint} \label{sec:constraint}

As a first step in understanding why $\Delta P|_\text{3-loop}$ can be subjected to an analytical 
treatment at high temperatures it is necessary to elucidate 
the properties of the generic constraint (in dimensionless quantities)
\[\label{eq:generic-constraint}
|(a+b)^2|=|2m^2-2\sqrt{\mathbf{a}^2+m^2}\sqrt{\mathbf{b}^2+m^2}-2|\mathbf{a}||\mathbf{b}|\cos \alpha|\leq1\,,
\]
where $\alpha\equiv\angl \mathbf{a}\mathbf{b}$. Recall from Sec.\,\ref{SEC:2} that the plus sign in $|(a+b)^2|\leq1$ implies $\sgn (a^0)=-\sgn (b^0)$, explaining the sign in the second term on the right hand side. For later use, let us consider two limits. For small spatial momenta, $|\mathbf{a}|,|\mathbf{b}|\ll1$, one has
\[
|(a+b)^2|\approx |2m^2-2\sqrt{m^2}\sqrt{m^2}|=0<1.
\]
That is, angle $\alpha$ becomes irrelevant. This can be made even more precise. It is straightforward to argue that
\[
\begin{split}\label{eq:genconstrest}
|(a+b)^2|&=|2m^2-2\sqrt{\mathbf{a}^2+m^2}\sqrt{\mathbf{b}^2+m^2}-2|\mathbf{a}||\mathbf{b}|\cos \alpha|\\
&\leq|2m^2-2\sqrt{\mathbf{a}^2+m^2}\sqrt{\mathbf{b}^2+m^2}-2|\mathbf{a}||\mathbf{b}||\,.
\end{split}
\]
For $|\mathbf{a}|\leq1/2$ and $|\mathbf{b}|\leq1/2$, the right hand side of \eqref{eq:genconstrest} can be estimated as
\[
\begin{split}
&|2m^2-2\sqrt{\mathbf{a}^2+m^2}\sqrt{\mathbf{b}^2+m^2}-2|\mathbf{a}||\mathbf{b}||\\\leq &\left|2m^2-2\left(\frac14+m^2\right)-2\cdot\frac14\right|=\left|-\frac12-\frac12\right|=1\,.
\end{split}
\]
As a consequence, \eqref{eq:generic-constraint} is satisfied independently of angle $\alpha$ if $|\mathbf{a}|\leq1/2$ and $|\mathbf{b}|\leq1/2$. However, this region yields a small contribution to $\Delta P|_\mathrm{3-loop}$, in particular at high temperatures.

\begin{figure}
{\centering
    \begin{tikzpicture}
      \begin{axis}[
          width=.7\linewidth, 
          xlabel=$|\mathbf b|/|\mathbf a|$, 
          ylabel=$\alpha$,
          ylabel style ={at={(-0.15,0.5)},anchor=south},
          minor x tick num={1},
          minor y tick num={1},
          xmin=0,
          ymin=0,ymax=3.14159,
          ytick={0,0.7854,...,10},
          scaled y ticks={real:3.1415},
          scaled ticks=false,
          yticklabels={$0$,${\pi}/{4}$,${\pi}/{2}$,${3\pi}/{2}$,$\pi$},
          legend style={at={(0.8,0.1)},anchor=south}, 
        ]
	\addplot[
        	smooth,
        	line width=2pt,
        ] 
        table[col sep=comma] {graphs/constraint-a-1.csv}; 
        \addlegendentry{$|\mathbf a|$=0.50}
        \addplot[
        	smooth,
        	dashed,
        	line width=2pt,
        ] 
        table[col sep=comma] {graphs/constraint-a-2.csv}; 
        \addlegendentry{$|\mathbf a|$=0.75}
        \addplot[
        	smooth,
        	dashdotted,
        	line width=2pt,
        ] 
        table[col sep=comma] {graphs/constraint-a-4.csv}; 
        \addlegendentry{$|\mathbf a|$=1.25}
        \addplot[
        	smooth,
        	dashdotdotted,
        	line width=2pt,
        ] 
        table[col sep=comma] {graphs/constraint-a-5.csv}; 
        \addlegendentry{$|\mathbf a|$=2.00}	
	\end{axis}
    \end{tikzpicture}
\hphantom{$0.999\pi$}
\\
    \begin{tikzpicture}
     	 \begin{axis}[
          width=.7\linewidth, 
          xlabel=$|\mathbf b|/|\mathbf a|$, 
          ylabel=$\alpha$,
          minor x tick num={1},
          minor y tick num={2},
          ymax=3.14159,
          ytick={3.11960,3.12274,3.12588,3.12903,3.13218,3.13531,3.13845,3.14159},
          scaled y ticks={real:3.1415},
          scaled ticks=false,
          yticklabels={$0.993\pi$,$0.994\pi$,$0.995\pi$,$0.996\pi$,$0.997\pi$,$0.998\pi$,$0.999\pi$,$\pi$},
          legend style={at={(0.8,0.1)},anchor=south}, 
        ]
	\addplot[
        	smooth,
        	line width=2pt,
        ] 
        table[col sep=comma] {graphs/constraint-b-1.csv}; 
        \addlegendentry{$|\mathbf a|$=50}
        \addplot[
        	smooth,
        	dashed,
        	line width=2pt,
        ] 
        table[col sep=comma] {graphs/constraint-b-2.csv}; 
        \addlegendentry{$|\mathbf a|$=100}
        \addplot[
        	smooth,
        	dashdotted,
        	line width=2pt,
        ] 
        table[col sep=comma] {graphs/constraint-b-3.csv}; 
        \addlegendentry{$|\mathbf a|$=300}
        \addplot[
        	smooth,
        	dashdotdotted,
        	line width=2pt,
        ] 
        table[col sep=comma] {graphs/constraint-b-4.csv}; 
        \addlegendentry{$|\mathbf a|$=1000}
	\end{axis}

    \end{tikzpicture}
    \hphantom{$0.999\pi$}
\caption{
Visualisation of $\alpha=\angl\mathbf{a}\mathbf{b}$ as a function of $|\mathbf{b}|/|\mathbf{a}|$ for small and large values of $|\mathbf{a}|$, obtained by saturating the constraint to $(a+b)^2=-1$ and setting the mass equal to the plateau value $m=2\sqrt{8}\pi$. The interpretation is that only coordinates above the lines are admitted by the constraint. When $|\mathbf{a}|$ surpasses 1, one approaches the situation described by \eqref{eq:deltaalpha} and \eqref{eq:deltab}, namely the width $\Delta\alpha$ approaches $|\mathbf{a}|^{-1}$ and the width $\Delta|\mathbf{b}|/|\mathbf{a}|$ approaches $0.113$ as $|\mathbf{a}|\rightarrow\infty$.
}
\label{fig:3-loop-ss}
}
\end{figure}
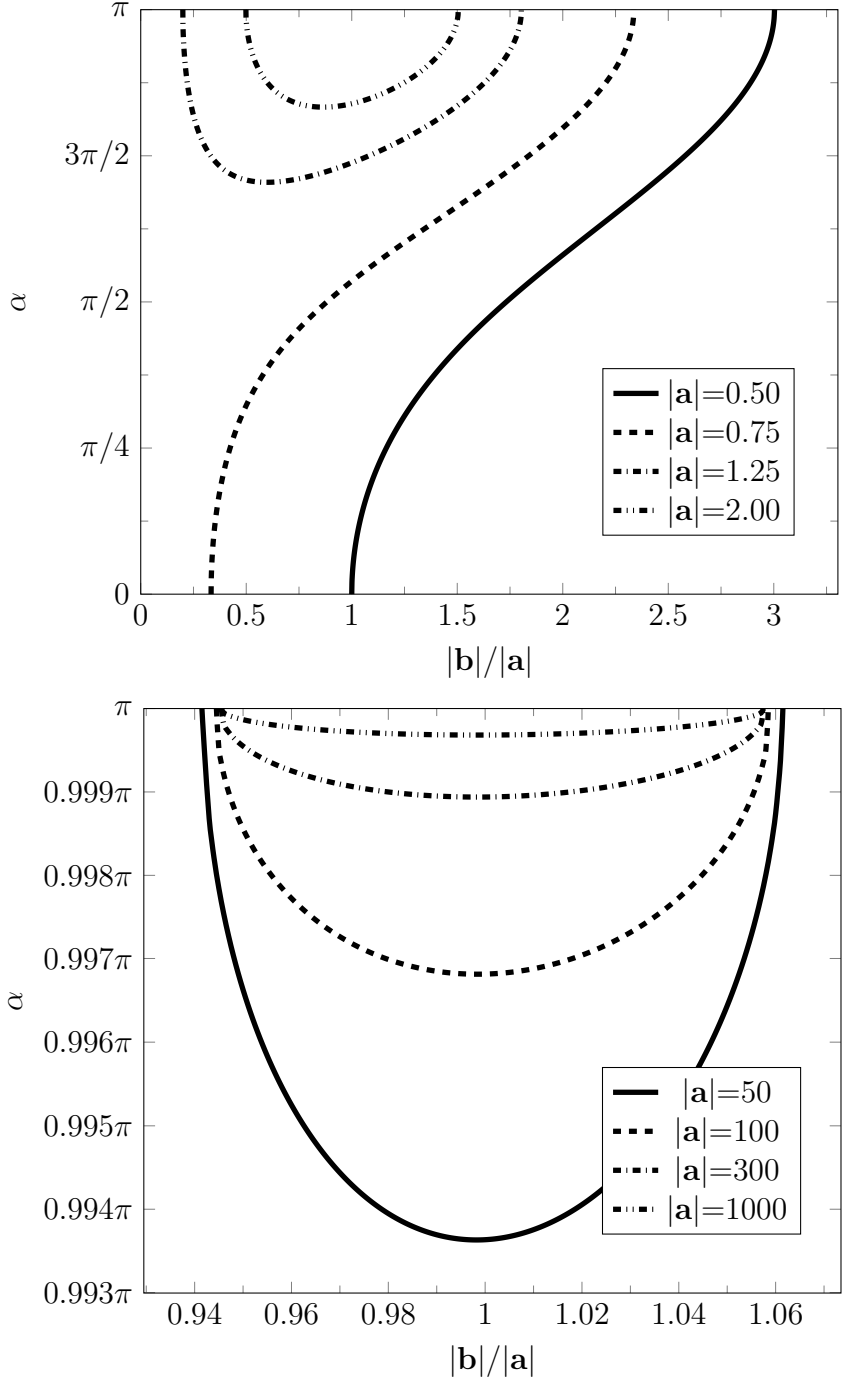

The second (and much more relevant) limit $|\mathbf{a}|,|\mathbf{b}|\gg1$ requires $\alpha$ to be close to $\pi$ and 
$|a|\approx|b|$. This is demonstrated by rendering the constraint \eqref{eq:generic-constraint} an equation by letting the right hand side vary as
\[
\label{eq:constraint-z}
(a+b)^2=-z,\quad z\in[-1,1]\,.
\] 
Solving \eqref{eq:constraint-z} for $\cos\alpha$ yields 
\[
\cos\alpha=\frac{z+2m^2-2\sqrt{\mathbf{a}^2+m^2}\sqrt{\mathbf{b}^2+m^2}}{2|\mathbf{a}||\mathbf{b}|}\,.
\]
For $|\mathbf{a}|,|\mathbf{b}|\gg m$ the condition $\cos\alpha\geq-1$ implies $z\geq0$, i.e. $z\in[0,1]$. 
Moreover, $\cos\alpha$ is maximised for fixed $|\mathbf{a}|$ and $z$ by
\[
|\mathbf{b}| = |\mathbf{a}|m\sqrt{\frac{4m^2-\frac{4m^2z}{|\mathbf{a}|^2}-\frac{z^2}{|\mathbf{a}|^2}}{4m^4+4m^2z+z^2}}\,,
\]
yielding, to leading order in $1/|\mathbf{a}|$,
\[
\cos\alpha\approx -1+\frac{1}{2|\mathbf{a}|^2}\left(z+\frac{z^2}{4m^2}\right) \,.
\]
Hence, $\cos\alpha$ is maximised by $z=1$ at fixed $|\mathbf{a}|$. The numerical value of the factor $1+1/(4m^2)$ is approximately $1.00079$ at high temperatures ($m=2e=2\sqrt{8}\pi$). This amounts to the width $\Delta\alpha$ of the band of allowed angles becoming narrower as $|\mathbf{a}|\rightarrow\infty$ like
\[\label{eq:deltaalpha}
\Delta\alpha=\pi- \arccos\left(-1+\frac{1}{2|\mathbf{a}|^2}\left(1+\frac{1}{4m^2}\right)\right) \approx \frac{1}{|\mathbf{a}|}\sqrt{1+\frac{1}{4m^2}} 
\]
to leading order in $1/|\mathbf{a}|$. Similarly, we derive the width $\Delta|\mathbf{b}|$ of the band of allowed values of $|\mathbf{b}|$ as follows: One can solve \eqref{eq:constraint-z} for $|\mathbf{b}|$, which is a quadratic equation and thus yields two values for $|\mathbf{b}|$. Their difference is maximised at $\cos\alpha=-1$. These two solutions at $\cos\alpha=-1$, 
referred to as $|\mathbf{b}|_\mathrm{max/min}$, are given by
\[\label{eq:solutionsb}
|\mathbf{b}|_\mathrm{max/min}=|\mathbf{a}|\left(1+\frac{z}{2m^2}\pm \frac12\sqrt{4\frac{z}{m^2}+\frac{z^2}{m^4}+\frac{4z}{|\mathbf{a}|^2}+\frac{z^2}{|\mathbf{a}|^2m^2} }\right)\,.
\]
Again, their difference is maximised by $z=1$, and variation of parameter $z\in[0,1]$ implies that all $|\mathbf{b}|$ between $|\mathbf{b}|_{\mathrm{max},z=1}$ and $|\mathbf{b}|_{\mathrm{min},z=1}$ are consistent with the constraint \eqref{eq:generic-constraint}.
At large $|\mathbf{a}|$ \eqref{eq:solutionsb} approaches
\[\label{eq:deltab}
|\mathbf{b}|_{\mathrm{max/min},z=1} = |\mathbf{a}|\left(1+\frac{1}{2m^2}\pm\frac12\sqrt{\frac4{m^2}+\frac1{m^4}}\right) 
\approx |\mathbf{a}|\left(1.002\pm0.056\right)
\]
where the expression to the far right, again, is obtained for the high-temperature 
value of $m$: $m=2\sqrt{8}\pi$. This means that allowed band of $|\mathbf{b}|$ values broadens in the limit $|\mathbf{a}|\rightarrow\infty$ and has a width of approximately \SI{11}{\%} of $|\mathbf{a}|$. The above limits for $\Delta\alpha$ and $|\mathbf{b}|_\mathrm{max/min}$ are rapidly approached, see \autoref{fig:3-loop-ss}. 

\subsection{High-temperature analysis}\label{sec:approximation}
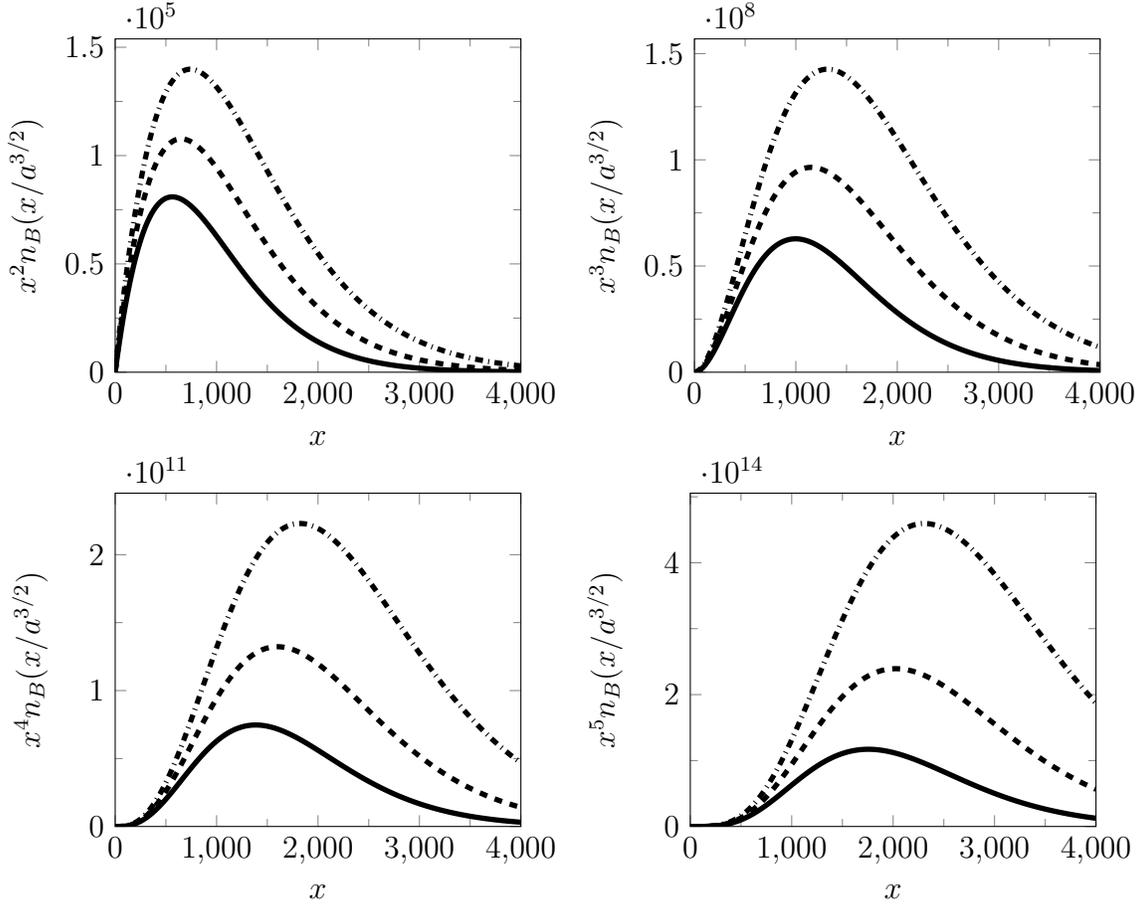
\begin{figure}
{\centering
    \begin{tikzpicture}
      \begin{axis}[
          width=.45\linewidth, 
          xlabel=$x$, 
          ylabel=$x^2n_B(x/a^{3/2})$,
          minor x tick num={1},
          minor y tick num={1},
          xmin=0,xmax=4000,
          ymin=10^-1,
        ]
      	\addplot[
	domain=0:4000,
	samples=201,
	line width=2pt,
	]
	{x^2*(exp(x /50^(3/2))-1)^(-1) };
	  \addplot[
	  dashed,
	domain=0:4000,
	samples=201,
	line width=2pt,
	]
	{x^2*(exp(x /55^(3/2))-1)^(-1) };
	  \addplot[
	dashdotted,
	domain=0:4000,
	samples=201,
	line width=2pt,
	]
	{x^2*(exp(x /60^(3/2))-1)^(-1) };
	\end{axis}
\end{tikzpicture}
\begin{tikzpicture}
 \begin{axis}[
          width=.45\linewidth, 
          xlabel=$x$, 
          ylabel=$x^3n_B(x/a^{3/2})$,
          minor x tick num={1},
          minor y tick num={1},
          xmin=0,xmax=4000,
          ymin=10^-1,
        ]
      	\addplot[
	domain=0:4000,
	samples=201,
	line width=2pt,
	]
	{x^3*(exp(x /50^(3/2))-1)^(-1) };
	  \addplot[
	  dashed,
	domain=0:4000,
	samples=201,
	line width=2pt,
	]
	{x^3*(exp(x /55^(3/2))-1)^(-1) };
	  \addplot[
	  dashdotted,
	domain=0:4000,
	samples=201,
	line width=2pt,
	]
	{x^3*(exp(x /60^(3/2))-1)^(-1) };
	\end{axis}
    \end{tikzpicture}

\begin{tikzpicture}
 \begin{axis}[
          width=0.45\linewidth, 
          xlabel=$x$, 
          ylabel=$\vphantom{x^2}x^4n_B(x/a^{3/2})$,
          minor x tick num={1},
          minor y tick num={1},
          y label style={at={(-.14,0.5)}},
          xmin=0,xmax=4000,
          ymin=10^-1,
          legend style={at={(1,-0.4)},anchor=east},
        ]
      	\addplot[
	domain=0:4000,
	samples=201,
	line width=2pt,
	]
	{x^4*(exp(x /50^(3/2))-1)^(-1) };
	  \addplot[
	  dashed,
	domain=0:4000,
	samples=201,
	line width=2pt,
	]
	{x^4*(exp(x /55^(3/2))-1)^(-1) };
	  \addplot[
	  dashdotted,
	domain=0:4000,
	samples=201,
	line width=2pt,
	]
	{x^4*(exp(x /60^(3/2))-1)^(-1) };
	\end{axis}
    \end{tikzpicture}
\begin{tikzpicture}
 \begin{axis}[
          width=0.45\linewidth, 
          xlabel=$x$, 
          ylabel=$\vphantom{x^3}x^5n_B(x/a^{3/2})$,
          minor x tick num={1},
          minor y tick num={1},
          y label style={at={(-.14,0.5)}},
          xmin=0,xmax=4000,
          ymin=10^-1,
          legend style={at={(1,-0.4)},anchor=east},
        ]
      	\addplot[
	domain=0:4000,
	samples=201,
	line width=2pt,
	]
	{x^5*(exp(x /50^(3/2))-1)^(-1) };
	  \addplot[
	  dashed,
	domain=0:4000,
	samples=201,
	line width=2pt,
	]
	{x^5*(exp(x /55^(3/2))-1)^(-1) };
	  \addplot[
	  dashdotted,
	domain=0:4000,
	samples=201,
	line width=2pt,
	]
	{x^5*(exp(x /60^(3/2))-1)^(-1) };
	\end{axis}
    \end{tikzpicture}
\caption{Plot of $f(x)=x^Nn_B(x/a^{3/2})$, for $N=2,3,4,5$ and $a=50$ (solid lines), $a=55$ (dashed lines), $a=60$ (dashdotted lines). 
This demonstrates that the qualitative features do not depend on $N\geq2$. At fixed $N$, maxima are shifted towards 
larger $x$ and $f(x)$ as $a$ increases. At fixed $a$, increasing $N$ also shifts the position of the maxima to larger $x$ and $f(x)$ increases by several orders of magnitude.}
\label{fig:polynomialtimesbose}
}
\end{figure}

Returning to the 3-loop integral \eqref{eq:3-loop-start}, we aim for an approximation in the high-temperature situation  $\lambda\gg\lambda_c$. Then the only temperature dependent entities within the integrand are the Bose factors since 
mass and coupling saturate to $m=2e=2\sqrt{8}\pi$. Let us exhibit why the limit $r_1,r_2\gg m$ is relevant. 
The variables $r_1$ and $r_2$ are integrated up to infinity but the factors of
\[
n_B\left(2\pi\sqrt{r_{1,2}^2+m^2}/\lambda^{3/2}\right)
\]
start to suppress the integrand when $2\pi(r_{1,2}^2+m^2)^{1/2}\gsim\lambda^{3/2}\log(2)$. The other factors in the integrand of \eqref{eq:3-loop-start} amount to a polynomial in radial (and other) variables, when considering this regime at high temperatures to approximate $(r_{1,2}^2+m^2)^{1/2}\approx r_{1,2}$. 
Hence, we expect terms proportional to $r_{1,2}^Mn_B'(r_{1,2})$ to be integrated over, where $M\leq N$ and $N$ is the (yet unknown) order of the polynomial. Since we are considering the limit of large $r_{1,2}$ the highest order term $r_{1,2}^Nn_B'(r_{1,2})$ should dominate. Regardless of the exact power $N$ (as long as it is greater than one), one obtains a curve similar to the Planck law, as shown in \autoref{fig:polynomialtimesbose} for several powers. The position of the maximum $r_{1,2;\mathrm{max}}$ is in all cases proportional to $\lambda^{3/2}$, and hence the integration is expected to be dominated by a small region centered at this maximum. This justifies to consider $r_1,r_2\gg m$. The 
diagonal contribution indeed is self-consistently given by this limit and subject to considerable simplifications.

From the discussion in Sec.\,\ref{sec:constraint} we know that for the $s$-channel constraint $|(p_1^2+p_4^2)|=|(p_2^2+p_3^2)|\leq1$ one has
 $\cos\sphericalangle\mathbf{p_1}\mathbf{p_4}\approx-1$ and $\cos\sphericalangle\mathbf{p_2}\mathbf{p_3}\approx-1$ and, using $\cos(\pi-\alpha)=-\cos\alpha$, we conclude
\[
\cos\sphericalangle\mathbf{p_1}\mathbf{p_2}=\cos\theta_1\approx-\cos\sphericalangle\mathbf{p_1}\mathbf{p_3}
\approx-\cos\sphericalangle\mathbf{p_2}\mathbf{p_4}\approx\cos\sphericalangle\mathbf{p_3}\mathbf{p_4}\,.
\]  
Exploiting these relations, \eqref{eq:r3} simplifies to
\begin{multline}
\sgn(p^0_2)\sqrt{r_2^2+m^2}+\sgn(p^0_3)\sqrt{r^2_3+m^2}-\sqrt{r_1^2+m^2}=
-\left[r^2_1+r^2_2+r^2_3 \right.\\\left. - 2 r_1r_2\cos\theta_1  +2r_1r_3\cos\theta_1-2r_2r_3 + m^2\right]^{1/2}\,.
\end{multline}
Recalling that $\sgn(p^0_2)=-\sgn(p^0_3)$ for all allowed sign configurations in the first sum of \eqref{eq:3-loop-start}, this is solved by $r_3=r_2$. Assuming $r_1,r_2>50$ and $r_3>0$, Mathematica confirms that this solution is unique. 
Thus it is legitimate in this limit to execute the sum over $r_3$ simply by replacing $r_3$ by $r_2$. Conveniently, this also implies $r_4=r_1$ which, in turn, entails that $p_1^0=-p^0_4$ and $p^0_2=-p^0_3$. Thus the two scalar products of four vectors reduce to
\begin{align}
p_1p_4&=p^0_1p^0_4-r_1r_4\cos\sphericalangle\mathbf{p_1}\mathbf{p_4} \approx -(p^0_1)^2 +r^2_1 = -m^2 \\
p_2p_3&= p^0_2p^0_3-r_2r_4\cos\sphericalangle\mathbf{p_2}\mathbf{p_3}\approx -(p^0_2)^2 +r^2_2 = -m^2\,,
\end{align}
and the remaining ones are expressed in terms of $p_1p_2$ as
\begin{align}
p_1p_2&= p^0_1p^0_2-r_1r_2\cos\sphericalangle\mathbf{p_1}\mathbf{p_2}\\
p_1p_3&\approx -p^0_1p^0_2+r_1r_2\cos\sphericalangle\mathbf{p_1}\mathbf{p_2}=-p_1p_2\\
p_2p_4&\approx -p^0_1p^0_2+r_1r_2\cos\sphericalangle\mathbf{p_1}\mathbf{p_2}=-p_1p_2\\
p_3p_4&\approx p^0_1p^0_2-r_1r_2\cos\sphericalangle\mathbf{p_1}\mathbf{p_2}=p_1p_2 \,.\\
\end{align}
Therefore, the only independent invariants in the polynomial in \eqref{eq:polynomial} are $m$ and 
\[
k^2(r_1,r_2,\theta_1)\equiv|p_1p_2|\,,
\]
and one can recast $P(\{p_i\})$ as 
\[
\begin{split}\label{eq:k}
&144-12\frac{1}{m^4}\left[2m^4+4k^4\right]+ 36\frac{1}{m^6}\left[4m^2k^4\right] +12\frac{1}{m^8}\left[k^8+m^8-2k^4m^4\right]\\
=&132+72\frac{k^4}{m^4}+12\frac{k^8}{m^8}\,
\end{split}
\]
which is strictly positive. The integrand of \eqref{eq:3-loop-start} reduces to
\[
r^2_1r^4_2\sin\theta_1\sin\theta_3\left(132+72\frac{k^4}{m^4}+12\frac{k^8}{m^8}\right)\frac{n_B'^2(r_1)n_B'^2(r_2)}{8(r_1^2+m^2)(r_2^2+m^2)},
\]
which, besides the factor $\sin\theta_3$, merely depends on $r_1,r_2$, and $\theta_1$. The $\phi_1$-integration can be executed trivially to yield a factor of $2\pi$. The factor of $\sin\theta_3$ appears to enforce the vanishing of the entire integral at first sight since we have assumed $\theta_3=\pi$. However, we know from the discussion in Sec.\,\ref{sec:constraint} that the width of the allowed deviation of $\theta_3$ from $\pi$ is given by 
\[
\Delta\theta_3 = \pi- \arccos\left(-1+\frac{1}{2r_2^2}\left(1+\frac{1}{4m^2}\right)\right) \approx \frac{1}{r_2}\sqrt{1+\frac{1}{4m^2}}\,.
\] 
Since this is increasingly small, the integration can be approximated by applying a mean value $\bar\theta_3=\pi-\Delta\theta_3/2$ as a function of $r_2$ and by multiplying this with the width $\Delta\theta_3$:
\[\label{eq:theta3integration}
\int_0^\pi\der\theta_3 \longrightarrow \int_0^\pi\der\theta_3\delta(\theta_3-\bar\theta_3)\Delta\theta_3\,.
\]
The full integral, subject to the remaining integrations over $r_1$, $r_2$, and $\theta_1$, thus reads
\begin{multline}\label{eq:approxint}
i\frac{\Lambda^4}{48\lambda^2}e^4\frac1{(2\pi)^5}\sum_\mathrm{signs}\intkern\der\theta_1\der r_1\der r_2r^2_1r^4_2\sin\theta_1\sin\left(\Delta\theta_3/2\right)\Delta\theta_3 \\
\times\left(132+72\frac{k^4}{m^4}+12\frac{k^8}{m^8}\right) 
\frac{n_B'^2(r_1)n_B'^2(r_2)}{8(r_1^2+m^2)(r_2^2+m^2)}\,
\end{multline}
where we have used that $\sin\left(\pi-x\right)=\sin(x)$. We now specialise to the diagonal contribution. (In the off-diagonal case the integration is further constrained by $|(p_1-p_2)^2| \leq1$.)

In the diagonal case, the integration over $\theta_1$ is unconstrained and thus for the first term of the residual polynomial \eqref{eq:k} the $\theta_1$-integral over $\mathrm{const.}\times\sin\theta_1$ amounts to a factor of $2$. For the second and third term consider that
\[
\int_0^\pi \der\theta \sin\theta\cos^n\theta
\]
vanishes for odd $n$. Therefore, all sign dependence in $k^4$ and $k^8$ drops out after integrating over $\theta_1$,
\[
\begin{split}
k^4&=(r_1^2+m^2)(r_2^2+m^2)+r_1^2r_2^2\cos^2\theta_1 \pm 2\sqrt{(r_1^2+m^2)(r_2^2+m^2)}r_1r_2\cos\theta_1,\\
      &=m^4+m^2(r_1^2+r_2^2)+r_1^2r_2^2(1+\cos^2\theta_1)  \\
      &\qquad\qquad\qquad\pm 2\sqrt{(r_1^2+m^2)(r_2^2+m^2)}r_1r_2\cos\theta_1 \\
k^8&=m^8+(r_1^2+r_2^2)2m^6 +r_1^2r_2^2m^4(4+6\cos^2\theta_1) + (r_1^4+r_2^4)m^4 \\
&\phantom{=}\quad (r_1^4r_2^2m^2+r_1^2r_2^4m^2) \cdot(2+6\cos^2\theta_1)  + r_1^4r_2^4(1+6\cos^2\theta_1+\cos^4\theta_1) \\
&\phantom{=}\quad \pm [\propto\cos^{(\mathrm{odd})}]\,.
\end{split}
\]
Hence $\theta_1$ can explicitly be integrated away, and, since the sign dependence is lost, the sum over signs simply amounts to a factor given by the number of possible energy-sign combinations. This factor is two in the diagonal case, see Sec.\,\ref{sec:3-loop-signs}. Upon processing all numerical factors in \eqref{eq:approxint} further, we arrive at 
\begin{multline}\label{eq:approxintss}
\frac{1}{3}\Delta P|_{\mathrm{3-loop},ss}=i\frac{\Lambda^4}{\lambda^2}e^4\frac1{(2\pi)^5}\intkern\der r_1\der r_2 \frac{r^2_1r^4_2}{(r^2_1+m^2)(r^2_2+m^2)}\sin\left(\Delta\theta_3/2\right)\Delta\theta_3\\
\times\left(\frac34+\frac{1}{3}\frac{r_1^2+r_2^2}{m^2}
+\frac7{12}\frac{r_1^2r_2^2}{m^4} +\frac1{24}\frac{r^4_1+r^4_2}{m^4}+ \frac1{6}\frac{r_1^4r_2^2+r_1^2r_2^4}{m^6} + \frac2{15}\frac{r_1^4r_2^4}{m^8}
\right) \\
\times n_B'^2(r_1)n_B'^2(r_2)\,,
\end{multline}
and we finally explicitly observe the anticipated polynomial structure in $r_1$ and $r_2$. Our assumption that the integrand is dominated by large $r_1$ and $r_2$ hence turns out to be self-consistent. Equation \eqref{eq:approxintss} represents the high-temperature limit of \eqref{eq:3-loop-start}, and the remaining integrals can be performed numerically. 
Before we treat the off-diagonal case, we demonstrate how the leading power in $\lambda$ can be 
extracted in \eqref{eq:approxintss}. Using 
\[
\Delta\theta\approx \frac{1}{r_2}\sqrt{1+\frac{1}{4m^2}}\, \quad \sin(\Delta\theta_3/2)\approx\frac1{2r_2}\sqrt{1+\frac{1}{4m^2}}
\]
simplifies \eqref{eq:approxintss} as 
\begin{multline}\label{eq:approxintss2}
\frac13\Delta P|_{\mathrm{3-loop},ss}=i\frac{\Lambda^4}{\lambda^2}e^4\frac1{(2\pi)^5}\left(1+\frac{1}{4m^2}\right){4m^2}\intkern\der r_1\der r_2 \frac{r^2_1r^2_2}{(r^2_1+m^2)(r^2_2+m^2)}\\
\times\left(\frac38+\frac16\frac{r_1^2+r_2^2}{m^2}
+\frac7{24}\frac{r_1^2r_2^2}{m^4} +\frac1{48}\frac{r^4_1+r^4_2}{m^4}+ \frac1{12}\frac{r_1^4r_2^2+r_1^2r_2^4}{m^6} + \frac1{15}\frac{r_1^4r_2^4}{m^8}
\right)\\
\times n_B'^2(p_1)n_B'^2(p_2)\,.
\end{multline}
Since the integrations over $r_1$ and $r_2$ in \eqref{eq:approxintss2} are independent from one another some factorisation 
subject to the following factors occurs
\[
\begin{split}
I_1&=\intkern\der x \frac{x^2}{x^2+m^2}n_B^2\left(2\pi\sqrt{\frac{x^2+m^2}{\lambda^{3}}}\right)\,,\\
I_2&=\intkern\der x \frac{x^4}{x^2+m^2}n_B^2\left(2\pi\sqrt{\frac{x^2+m^2}{\lambda^{3}}}\right)\,,\\
I_3&=\intkern\der x \frac{x^6}{x^2+m^2}n_B^2\left(2\pi\sqrt{\frac{x^2+m^2}{\lambda^{3}}}\right)\,.
\end{split}
\]
Although only $I_3$ turns out to be relevant for the leading order in $\lambda$ let us analyse all three integrals in a 
high-temperature situation. Namely, for sufficiently large $\lambda$ one can split $I_1$ into a first part $I^{\hat x}_{1}$, integrated up to an arbitrary parameter $\hat x$ such that $m\ll\hat x\ll\lambda^{3/2}\log(2)/(2\pi)$ and a second part $I^{\infty}_{1}$ from $\hat x$ to infinity. For simplicity, we choose $\hat x = \lambda$. For the integration in $I^{\hat x}_{1}$ we may apply
\[\label{eq:boseapprox}
n_B\left(2\pi\sqrt{\frac{x^2+m^2}{\lambda^3}}\right)\approx \frac{1}{1+2\pi\sqrt{\frac{x^2+m^2}{\lambda^3}}-1} = \frac1{2\pi}\sqrt{\frac{\lambda^3}{(x^2+m^2)}}\,.
\]
Therefore, $I^{\hat x}_1$ can be approximated as
\[
\begin{split}\label{eq:i1a}
I^{\hat x}_1&\approx\frac{\lambda^3}{4\pi^2}\int_{0}^{\hat x}\der x\frac{x^2}{(x^2+m^2)^2} = \frac{\lambda^3}{4\pi^2m}\int_{0}^{\hat y}\der y\frac{y^2}{(y^2+1)^2} \\
&= \frac{\lambda^3}{4\pi^2m}\frac12\left(\left[-\frac{y}{y^2+1}\right]^{\hat{y}}_{0} + \int_{0}^{\hat y}\der y\frac{1}{y^2+1}\right) \\
&= \frac{\lambda^3}{8\pi^2}\left(-\frac{\hat x}{\hat x^2+m^2} + \frac{\arctan\left(\frac{\hat x}{m}\right)}{m}\right)
, \\
&\approx \frac{\lambda^3}{16\pi m}-\frac{\lambda^2}{8\pi^2}\,,
\end{split}
\]
where $y\equiv x/m$, and $\arctan\left(\frac{\hat x}{m}\right)\approx\pi/2$. 
For the remaining integral $I^\infty_1$ the prefactor of $n_B^2$ in the integrand is close to unity, and we obtain
\[
\begin{split}\label{eq:i1b}
I^\infty_1&\approx\int_{\hat x}^\infty\der x\, n_B^2\left(2\pi\sqrt{\frac{x^2+m^2}{\lambda^3}}\right) 
= \frac{\lambda^{3/2}}{2\pi}\int_{\hat y}^\infty\der y\, n_B^2\left(y\right) \\
&= \frac{\lambda^{3/2}}{2\pi}\left[-\hat y + n_B(\hat y) +\log(e^{\hat y}-1)\right] \\
&\approx -\lambda + \frac{\lambda^2}{4\pi^2} - \frac{\lambda^{3/2}}{2\pi}\log\left(\frac{\lambda^{1/2}}{2\pi}\right)\,,
\end{split}
\]
where $y\equiv2\pi x/\lambda^{3/2}$. Thus, indeed, the contribution of leading power three in $\lambda$ resides in $I^{\hat x}_1$, and we conclude
\[
I_1\approx \frac{\lambda^3}{16\pi m}
\]
for sufficiently high $\lambda$. $I_2$ and $I_3$ are, due to the higher powers in the prefactors 
of $n_B^2$ in the respective integrands, strongly dominated by $x^2\gg m^2$ and thus we can approximate
\begin{equation*}
I_2\approx \int_0^{\infty}\der x \, x^2n_B^2\left(2\pi\sqrt{\frac{x^2}{\lambda^{3}}}\right) 
= \frac{\lambda^{9/2}}{(2\pi)^3}\int_0^\infty\der y\,  y^2n_B^2(y)
= \frac{\lambda^{9/2}}{(2\pi)^3\cdot3}(\pi^2-6\zeta(3))
\end{equation*}
and 
\begin{equation*}
I_3\approx \int_0^{\infty}\der x \, x^4n_B^2\left(2\pi\sqrt{\frac{x^2}{\lambda^{3}}}\right) 
= \frac{\lambda^{15/2}}{(2\pi)^5}\int_0^\infty\der y\,  y^4n_B^2(y)
= \frac{\lambda^{15/2}}{(2\pi)^5}\frac{4}{15}(\pi^4-90\zeta(5))
\end{equation*}
where $\zeta(z)$ denotes Riemann's zeta function, and we have substituted 
$y=2\pi x/\lambda^{3/2}$. Thus, to leading order in $\lambda$ we have 
\[
\begin{split}\label{eq:approxintss3}
\frac{1}{3}\Delta P|_\mathrm{\mathrm{3-loop},ss} &= i\frac{\Lambda^4}{\lambda^2}e^4\frac1{(2\pi)^5}\frac1{15}\left(1+\frac{1}{4m^2}\right)\frac{I_3^2}{m^8}\\
&=i\Lambda^4\frac{1}{3375}\frac{1}{(2\pi)^{15}}\frac{1}{m^4}\left(1+\frac{1}{4m^2}\right)\left(\pi^4-90\zeta(5)\right)^2\lambda^{13}\\ 
&\equiv ic_{13}\Lambda^4\lambda^{13}
\end{split}
\]
for $\lambda\gg\lambda_c$. In \eqref{eq:approxintss3} $m=2e$ was used. The numerical value of coefficient $c_{13}$ is 
\[
c_{13}=5.2968\cdot10^{-20}\,.
\] 
This value is confirmed by a fit to the numerical integration of \eqref{eq:approxintss}.

Returning to the off-diagonal case, one has to assert $|(p_1-p_2)^2|\leq1$ in \eqref{eq:approxint}. In this case the limit $r_1,r_2\gg1$ implies that $\theta_1\approx 0$ instead of $\pi$, and one can estimate the width and mean value 
of the allowed band for $\theta_1$ in the same way as for $\theta_3$ in \eqref{eq:theta3integration}. This yields 
\begin{align}
\Delta\theta_1&=\arccos\left(1-\frac1{2r_1r_2}\sqrt{1+\frac{1}{4m^2}}\right)\approx\arccos\left(1-\frac1{2r_1r_2}\right)\,.
\end{align}
However, the constraint also implies a limited width of $r_1$, depending on $r_2$. The maximum width $\Delta r_1$ and mean $\bar r_1$ were calculated in Sec.\,\ref{sec:constraint} to be
\[
\Delta r_1 = r_2\sqrt{\frac{4}{m^2}+\frac1{m^4}}, \qquad \bar r_1= r_2\left(1+\frac1{2m^2}\right)\,.
\]
This can be used to replace the $\theta_1$ and $r_1$ integrations in \eqref{eq:approxint} in analogy to the case of $\theta_3$ in \eqref{eq:theta3integration}, and one obtains
\begin{multline}\label{eq:approxintst}
\frac23\Delta P|_{\mathrm{3-loop},st} = i\frac{\Lambda^4}{576\lambda^2}e^4\frac1{(2\pi)^5}
\intkern\der r_2\frac{r^2_1r^4_2}{(r_1^2+m^2)(r_2^2+m^2)}\Delta r_1\sin\left(\Delta\theta_1/2\right)\Delta\theta_1
\\\times\sin\left(\Delta\theta_3/2\right)\Delta\theta_3 
\left(132+72\frac{k^4}{m^4}+12\frac{k^8}{m^8}\right) 
{n_B'^2(r_1)n_B'^2(r_2)}\,.
\end{multline}
In \eqref{eq:approxintst} $k$ is evaluated at $\theta_1=\Delta\theta_1/2,$ $r_1=r_2(1+1/{2m^2})$, and the sum over energy-sign combinations has been dropped noting that $\sgn(p^0_1)=\sgn (p^0_2)$. This expression can be evaluated numerically.
Working to first order in $1/r_2$, as in the diagonal case, however, renders the polynomial a constant since
\[
k^4=(p_1p_2)^2\approx(r_1r_2(1-\cos\bar\theta_1))^2\approx r_1^2r_2^2\left(\frac12\cdot\frac1{(2\sqrt{r_1r_2})^2}\right)^2 = \frac1{64}\,.
\]
Approximating the mean $\bar r_1=r_2(1+1/(2m^2))$ by $r_2$, one finally obtains
\[\label{eq:approxintst2}
\frac23\Delta P|_{{3-loop},st} \approx i\frac{\Lambda^4}{\lambda^2}e^4 C \intkern\der r_2\frac{r_2^3}{(r_2^2+m^2)^2}n_B^4\left(2\pi\sqrt{\frac{r_2^2+m^2}{\lambda^3}}\right)\,,
\]
where we have introduced the constant
\[
\begin{split}
C&\equiv\frac{1}{2304}\frac{1}{(2\pi)^5}\sqrt{\frac4{m^2}+\frac1{m^4}}\left(132+\frac{72}{64}\frac1{m^4}+\frac{3}{1024}\frac{1}{m^8}\right) \\
& \approx 6.5867\cdot10^{-7}\,.
\end{split}
\]
This integral is dominated by $r_2\ll\lambda$. As was done for $I_1$ in \eqref{eq:i1a} and \eqref{eq:i1b}, we aim for a split of the integration 
\[
I_4\equiv\int_0^\infty\der x\frac{x^3}{(x^2+m^2)^2}n^4_B\left(2\pi\sqrt{\frac{x^2+m^2}{\lambda^3}}\right)\,.
\]
To be able to use \eqref{eq:boseapprox} in one domain and the approximation
\[\label{eq:boseapprox2}
n_B(x)=\frac{1}{e^x-1}\approx e^{-x}\,,
\]
reliable for large $x$, in the other domain, we choose to split the integration at $\hat x=\lambda^{3/2}/(2\pi)$ where the exponential becomes unity upon neglecting $m$. Using \eqref{eq:boseapprox}, we then find
\[
\begin{split}
I_4^{\hat x} &= \int_0^{\hat x}\der x\frac{x^3}{(x^2+m^2)^2}n^4_B\left(2\pi\sqrt{\frac{x^2+m^2}{\lambda^3}}\right) 
\approx \frac{\lambda^6}{(2\pi)^4}\int_0^{\hat x}\der x\frac{x^3}{(x^2+m^2)^4} \\
&=\frac{\lambda^6}{(2\pi)^4}\frac{\hat x^6+ 3m^2 \hat x^4 }{12 m^4 (m^2 +\hat x^2)^3}\\
&\approx\frac{\lambda^6}{(2\pi)^4}\frac{1}{12m^4(1+\frac{m^2}{\hat x^2})^3}\approx\frac{\lambda^6}{(2\pi)^4}\frac{1}{12m^4} \,
\end{split}
\]
where in the final line we omitted terms of lower order in $\lambda$. Notice that this leading order is independent of the precise upper boundary $\hat x$: For any $\hat x\propto\lambda^k$ with $k>0$ the first correction is proportional to $\lambda^6m^2/\hat x^2\propto\lambda^{6-2k}$, and thus it is subleading. This already indicates the saturation of $I_1$ well below $\hat{x}=\lambda^{3/2}/(2\pi)$. The other contribution to $I_4$, $I^\infty_4$ is estimated by virtue of \eqref{eq:boseapprox2} as
\[
\begin{split}
I_4^{\infty}&=  \int_{\hat x}^{\infty}\der x\frac{x^3}{(x^2+m^2)^2}n^4_B\left(2\pi\sqrt{\frac{x^2+m^2}{\lambda^3}}\right) 
\approx  \int_{\hat x}^{\infty}\der x\frac{1}{x}\left(e^{-2\pi\frac{x}{\lambda^{3/2}}}\right)^{4} \\
&= \int_1^\infty \der y \,y^{-1}e^{-4y} = \int_4^\infty \der \bar{y} \,\bar{y}^{-1}e^{-\bar{y}} = \Gamma(0,4)
\end{split}
\]
where, again, $y\equiv 2\pi x/\lambda^{3/2}$, $\bar{y}=4y$, and $\Gamma(s,t)$ denotes the incomplete gamma function. We notice 
that $I_4^{\infty}$ is constant in $\lambda$, and thus it can be dropped at high $\lambda$. 
Hence, we can recast \eqref{eq:approxintst2} as
\[
\begin{split}\label{eq:approxintst3}
\frac{2}{3}\Delta P|_{\mathrm{3-loop},st} &\approx  i\Lambda^4e^4\frac{C}{(2\pi)^4}\frac{1}{12m^4}\lambda^4\approx i\Lambda^4 \lambda^4\cdot2.2011\cdot 10^{-12}\,.
\end{split}
\]
The leading power in $\lambda$ thus is identified to be four. 
We conclude that, concerning contributions to \eqref{eq:approxint} from the integration range of large $r_1$ and $r_2$, the diagonal channel's contribution is leading by a relative power of $\lambda^{13-4}=\lambda^9$ in the high-temperature limit. Possible corrections from the integration domain of small radii can be accounted for in the Monte Carlo calculation and be compared to the approximation \eqref{eq:approxintst3}.

\subsection{Match of high- to low-temperature behaviour}\label{sec:results}
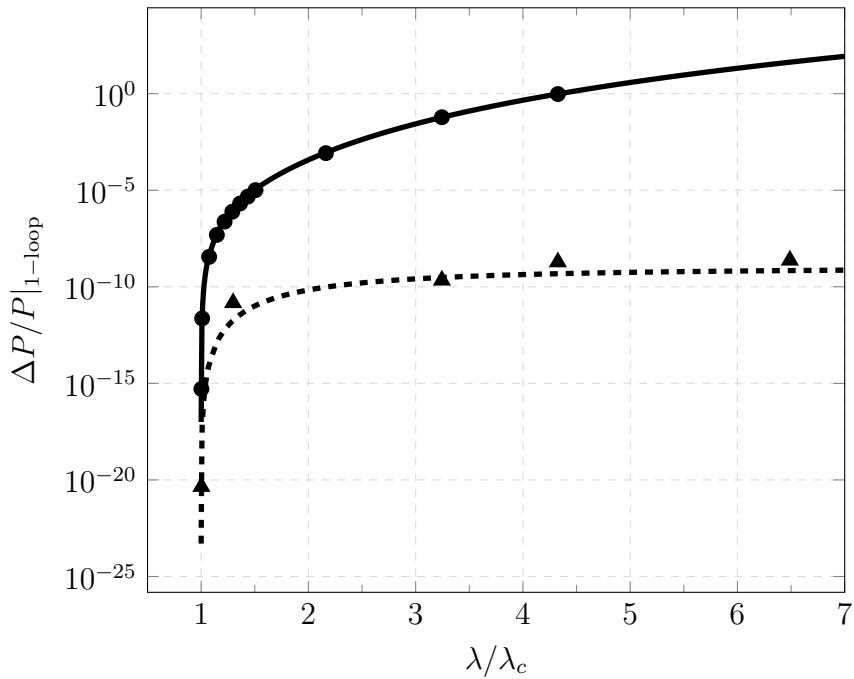
\begin{figure}
{\centering
    \begin{tikzpicture}
      \begin{semilogyaxis}[
          width=.7\linewidth, 
          grid=major, 
          grid style={dashed,gray!30}, 
          xlabel=$\lambda/\lambda_c$, 
          ylabel=$\Delta P/P|_{\mathrm{1-loop}}$,
          minor x tick num={1},
          minor y tick num={4},
          xmin=0.5, xmax=7,
        ]
        \addplot[
        line width=2pt,
        ] 
        table{graphs/3-loop-ss-results-approx.dat}; 
        
         \addplot[
         only marks,
        line width=2pt,
        ] 
        table{graphs/3-loop-ss-results-mc.dat}; 

        \addplot[
        smooth,
        dashed,
        line width=2pt,
        ] 
        table{graphs/3-loop-st-results-approx.dat}; 
        
        \addplot[
         only marks,
        line width=2pt,
         mark=triangle,
        ] 
        table{graphs/3-loop-st-results-mc.dat}; 
      \end{semilogyaxis}
    \end{tikzpicture}
    \hphantom{$10^{-25}$}
\caption{The moduli of the three-loop pressure corrections, $1/3\Delta P|_{\mathrm{3-loop},ss}$ (solid) and 
$2/3\Delta P|_{\mathrm{3-loop},st}$ (dashed), divided by the massive sector one-loop pressure $P|_{\mathrm{1-loop}}$. The continuous curves represent the analytical high-$T$ expressions \eqref{eq:approxintss} and \eqref{eq:approxintst} while the dots and triangles are the respective $ss$ and $st$ Monte Carlo (MC) results. The MC results for the case $st$ are less precise than those for $ss$ at comparable running time because the additional constraint in the $st$ case strongly decreases the MC hit rate.}
\label{fig:3-loop-results}}
\end{figure}

In \autoref{fig:3-loop-results} results for the 2PI three-loop pressure corrections $1/3\Delta P|_{\mathrm{3-loop},ss}$ and $2/3\Delta P|_{\mathrm{3-loop},st}$ are shown. (Recall that they add up to the full expression $\Delta P|_{\mathrm{3-loop}}$ in \eqref{eq:3-loop-start}.) We compare Monte Carlo evaluations at low temperatures with the approximations derived in Sec.\,\ref{sec:approximation}. The approximations are successful even for low values of $\lambda$ in the vicinity of $\lambda_c$. With \eqref{eq:approxintss} we thus have extracted a simple and extremely well approximating expression for the 2PI three-loop which rapidly saturates into the power law of \eqref{eq:approxintss3} with increasing $\lambda$. 

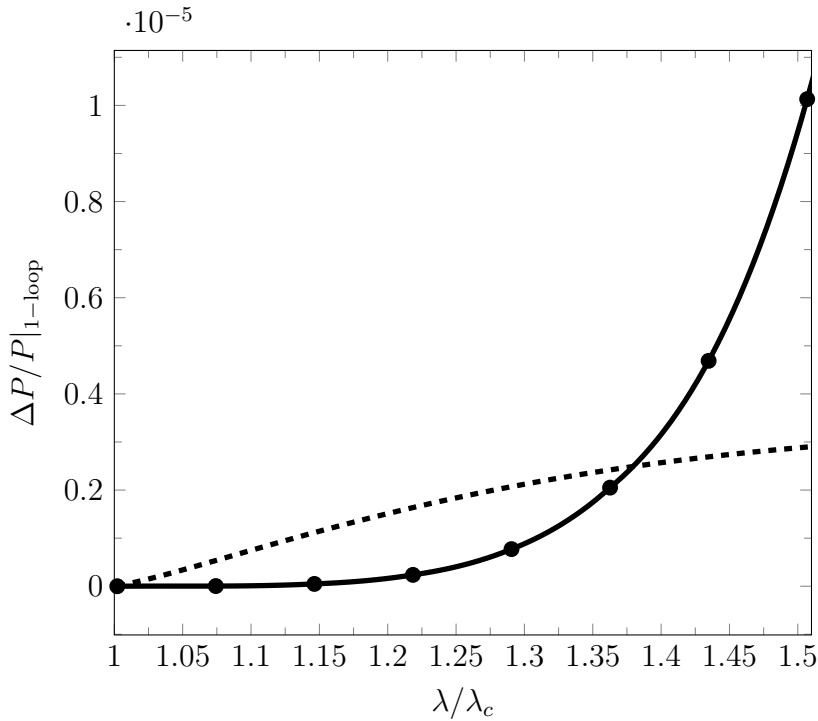
\begin{figure}
{\centering
    \begin{tikzpicture}
      \begin{axis}[
          width=.7\linewidth, 
          xlabel=$\lambda/\lambda_c$, 
          ylabel=$\Delta P/P|_{\mathrm{1-loop}}$,
          minor x tick num={1},
          minor y tick num={1},
          xmin=1, xmax=1.51,
        ]
        \addplot[
        dashed,
        smooth,
        line width=2pt,
        ] 
        table{graphs/2l-transition.dat}; 
        
         \addplot[
         only marks,
        line width=2pt,
        ] 
        table{graphs/3l-transition.dat}; 

        \addplot[
        smooth,
        line width=2pt,
        ] 
        table{graphs/3l-transition-interpolation.dat}; 
      \end{axis}
    \end{tikzpicture}
    \hphantom{$0.6$}
\caption{Monte Carlo results of $\Delta P|_{\mathrm{3-loop}}/P|_{\mathrm{1-loop}}$ close to $\lambda_c$ (dots). The solid line is a smooth interpolation of the latter while the dashed line represents $\Delta P|_{\mathrm{2-loop}}/P|_{\mathrm{1-loop}}$. 
}
\label{fig:3l-vs-2l}
}
\end{figure}

Compared to the two-loop contribution $\Delta P|_{\mathrm{2-loop}}$ in \autoref{fig:2-loop-plot}, whose ratio with the one-loop pressure $P|_\mathrm{1-loop}$ approaches zero at high temperatures, and compared to $P|_\mathrm{1-loop}$ in \eqref{eq:oneloop-pressuremassive} itself, which increases with a Stefan-Boltzmann power of four, the 2PI three-loop diagram thus dominates at sufficiently high temperatures. However, in the vicinity of $\lambda_c$, $|\Delta P|_{\mathrm{3-loop}}|$ is found by Monte Carlo integration to be well suppressed with respect to $P|_\mathrm{1-loop}$ and $|\Delta P|_{\mathrm{2-loop}}|$. The transition regime, where the modulus of $\Delta P|_{\mathrm{3-loop}}$ becomes larger than the modulus of $\Delta P|_{\mathrm{2-loop}}$, 
is shown in \autoref{fig:3l-vs-2l}.
Notice that both, $1/3\Delta P|_{\mathrm{3-loop},ss}$ and 
$2/3\Delta P|_{\mathrm{3-loop},st}$, are {\sl purely imaginary}, see \eqref{eq:approxintss} and \eqref{eq:approxintst}. 
This indicates that $\Delta P|_{\mathrm{3-loop}}$ radiative corrections to the free quasiparticle pressure cannot in general be interpreted as thermal. Rather, minute imaginary contributions, see Sec.\,\ref{SEC:5}, 
represent turbulent contributions to the pressure. This is not too surprising since they describe 
(minute) correlations between events induced by (anti)caloron centers, and we know that such correlations induce isolated, screened monopole-antimonopole pairs \cite{KellerRH2008spatialWilsonloop} whose condensation at low temperatures 
implies a highly nonthermal Hagedorn transition \cite{Hofmann2016}.     
 
As one would expect, the least-constrained channel combination by far dominates the 2PI three-loop pressure correction. Moreover, we have found the high-temperature limit of this contribution, namely \eqref{eq:approxintss} and its power law limit \eqref{eq:approxintss3}. The subleading contribution \eqref{eq:approxintst} is bounded relative to $P|_\mathrm{1-loop}$ since both are $\propto\lambda^4$ as $\lambda\rightarrow\infty$.

The unboundedness with power thirteen of \eqref{eq:approxintss3} as $\lambda\to\infty$ calls for a reorganisation of the loop expansion. In Sec.\,\ref{sec:n-gon-theorem} we have shown that a class ${\cal C}$ of diagrams with up to infinitely many loops exists where the loop integration is subject to a single constraint only just like in the diagonal configuration of the three-loop diagram. 
Resummation within ${\cal C}$ actually resolves the unboundedness problem of the fixed loop order three 
as we will show in this section. Yet, we emphasise that comparing \eqref{eq:approxintss3} and \eqref{eq:approxintst3} demonstrates the suppressing strength of vertex constraints: The imposition of an additional constraint leads to a 
hierarchical drop in the scaling power at high temperatures. Namely, a power law $\lambda^{13}$ 
reduces to a power law $\lambda^4$. This provides a certain amount of evidence that the consideration of the least constrained combinations are likely sufficient to identify contributions which make sense in a resummed form only.

\section{\label{SEC:5}Resummation of two-particle irreducible diagrams with dihedral symmetry} 
The Dyson-Schwinger equation for the resummed 4-vertex in scalar $\lambda\varphi^4$-theory without 
sources is given as \cite{swanson}
\begin{equation}\label{eq:dsphi4}
\includegraphics[scale=0.6,align=c]{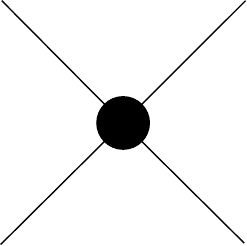} 
= \includegraphics[scale=0.6,align=c]{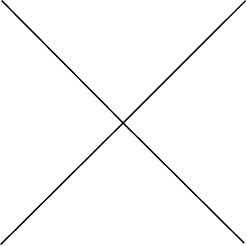} 
+ \includegraphics[scale=0.6,align=c]{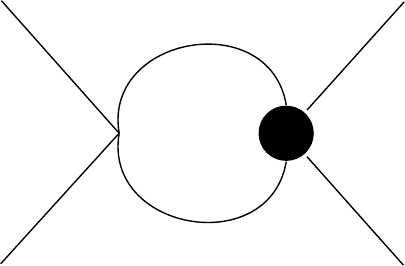}
+\includegraphics[scale=0.6,align=c]{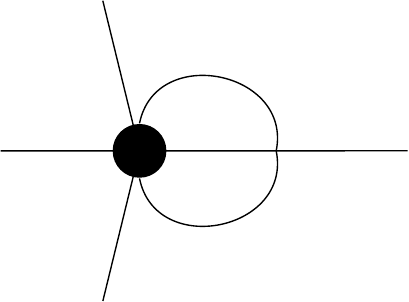}
+\includegraphics[scale=0.6,align=c]{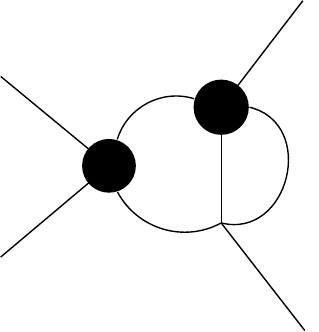}\,,
\end{equation}
where dotted vertices are exact (subject to all-order resummations), internal lines denote resummed propagators, while external lines refer to arbitrary asymptotic states. There are also tree-level 4-vertices in the diagrams on the right hand side. It is required by the SU(2) algebra su(2) that only those two massive particles, which are oppositely charged w.r.t. the unbroken U(1) Cartan subalgebra u(1), participate \cite{Hofmann2016}. 
Considering this restriction, one can take over the diagrammatic equation from $\lambda\varphi^4$ theory. We now 
demonstrate that \eqref{eq:dsphi4} can be used to generate a well bounded loop expansion of the pressure by resummation. 
To this end, let us approximate the full propagator by its tree-level expression while assuming that the resummed 4-vertex $\Gamma_{[4]}$ has the same tensorial structure as the tree-level 4-vertex, i.e. 
\[
\Gamma_{[4],abcd}^{\mu\nu\rho\sigma}|_i = f(\lambda,i)\Gamma_{[4],abcd}^{\mu\nu\rho\sigma}|_{i,\mathrm{tree-level}}
\]
where $i=s,t,u$, and $f(\lambda,i)$ is a scalar function (form factor) encoding the dependences on invariant momentum transfer and 
temperature. Let us further truncate the DS equation after the second diagram on the right hand side 
of \eqref{eq:dsphi4}. This step cannot at present be justified rigorously but it appears 
plausible that, due to reduced symmetry, the third and fourth diagrams on the right hand side of 
\eqref{eq:dsphi4} are much more constrained than the second one is. 
Under these assumptions, we may fix the color structure of this latter diagram since of the three options
\[\label{eq:dstensor}
\includegraphics[scale=.9,align=c]{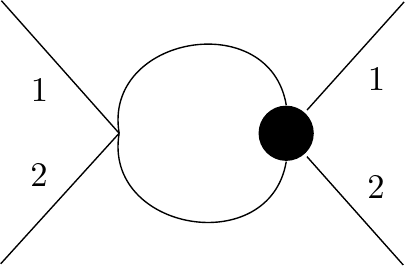}\,,\quad
\includegraphics[scale=.9,align=c]{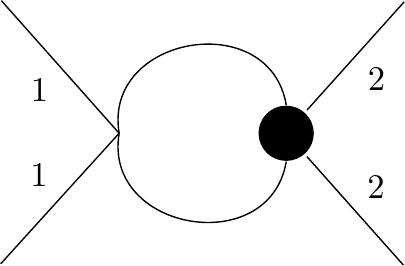}\,,\quad
\includegraphics[scale=.9,align=c]{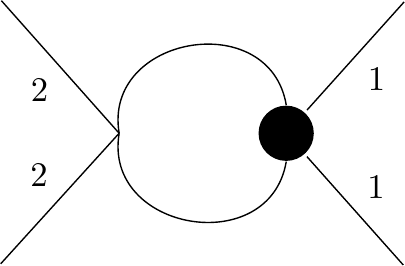}\,,
\]
where the 1 and 2 denote the respective massive directions in su(2), only the first one is compatible with the vertex color structure. In this truncation, one can iterate the DS equation starting at tree-level on the right hand side. At each iteration step one inserts the resummed vertex of the previous iteration as
\[\label{eq:iteration}
\begin{split}
\includegraphics[scale=0.5,align=c]{feynman-graphs/DS-4-vertex-1}_{\,1}& 
= \includegraphics[scale=0.5,align=c]{feynman-graphs/DS-4-vertex-2}  + \includegraphics[scale=0.5,align=c]{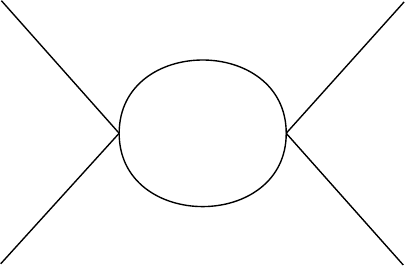} \\
\includegraphics[scale=0.5,align=c]{feynman-graphs/DS-4-vertex-1}_{\,2}&
=\includegraphics[scale=0.5,align=c]{feynman-graphs/DS-4-vertex-2}  + \includegraphics[scale=0.5,align=c]{feynman-graphs/DS-4-vertex-3}_{\,1}
=\includegraphics[scale=0.5,align=c]{feynman-graphs/DS-4-vertex-2}  + \includegraphics[scale=0.5,align=c]{feynman-graphs/DS-preiteration-1l} +\includegraphics[scale=0.5,align=c]{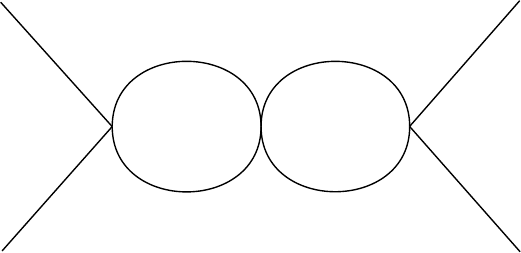}  \\
\includegraphics[scale=0.5,align=c]{feynman-graphs/DS-4-vertex-1}_{\,3}& 
=\includegraphics[scale=0.5,align=c]{feynman-graphs/DS-4-vertex-2}  + \includegraphics[scale=0.5,align=c]{feynman-graphs/DS-preiteration-1l} +\includegraphics[scale=0.5,align=c]{feynman-graphs/DS-preiteration-2l}+\includegraphics[scale=0.5,align=c]{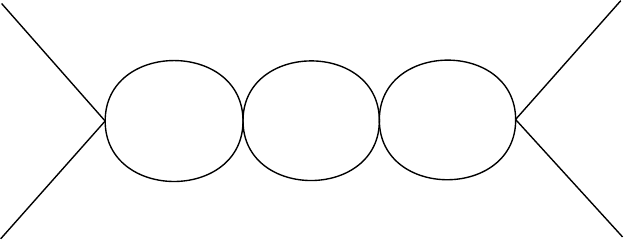}\\
&\dots
\end{split}
\]
where the numbers in subscript denote the level of iteration at which the resummed vertex is evaluated. Therefore, the full resummation of the class of diagrams shown in \eqref{eq:iteration} is encoded in the truncated version of the DS equation \eqref{eq:dsphi4}.
Closing the legs into two (extra) loops to compare with two-loop and three-loop calculations transforms the truncated version of \eqref{eq:dsphi4} into
\begin{equation}
\includegraphics[scale=0.8,align=c]{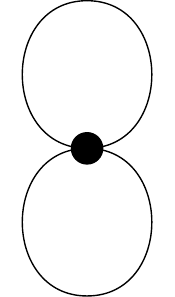} 
=\includegraphics[scale=0.8,align=c]{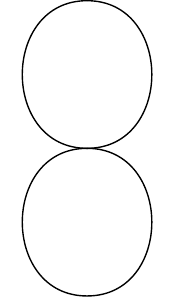} 
+\includegraphics[scale=0.8,align=c]{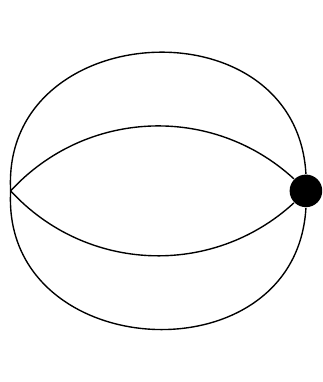}\,.
\end{equation}
Comparing with \eqref{eq:iteration}, this equation can be interpreted as the resummation of the class $\mathcal{C}$ as defined in Sec.\,\ref{sec:n-gon-theorem} since by iteration the chain
\[
\includegraphics[scale=0.6,align=c]{feynman-graphs/DS-closed-1} =
\includegraphics[scale=0.6,align=c]{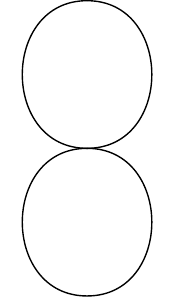}+
\includegraphics[scale=0.6,align=c]{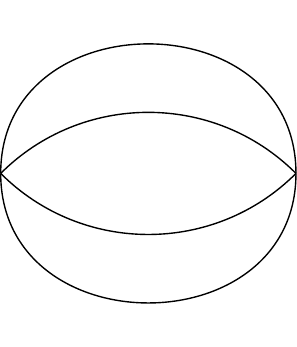}+
\includegraphics[scale=0.6,align=c]{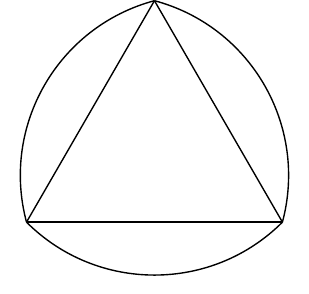}+
\includegraphics[scale=0.6,align=c]{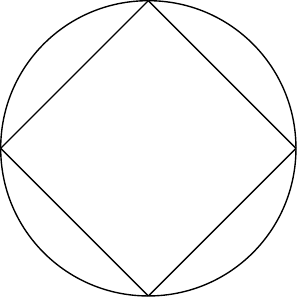}+
\includegraphics[scale=0.47,align=c]{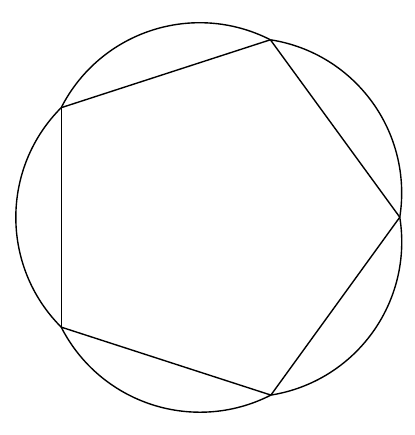}+\,\dots
\]
is generated. 
In the high-temperature limit, $f(\lambda,i)$ can be considered constant in 
momentum transfer, $f(\lambda,i)\equiv f(\lambda)$ since the constraints demand that
\[
|s|,|t|,|u|\leq|\phi|^2=\frac{\Lambda^3}{2\pi T}\propto 1/T \,.
\]
Thus $f(\lambda)$ can be factored out in loop integrals, and the corresponding pressure contributions $\Delta P|_\mathrm{2-loop}$ and $\Delta P|_\mathrm{3-loop}$ imply
\[
f(\lambda)\Delta P|_\mathrm{2-loop} = \Delta P|_\mathrm{2-loop} + f(\lambda)\Delta P|_\mathrm{3-loop}\,.
\]
Solving for $f(\lambda)$ one arrives at
\[\label{eq:foflambda}
f(\lambda) = \frac{\Delta P|_\mathrm{2-loop}}{\Delta P|_\mathrm{2-loop} - \Delta P|_\mathrm{3-loop}}\,.
\]
The fact that $\Delta P|_\mathrm{3-loop}$ is imaginary while $\Delta P|_\mathrm{2-loop}$ is real ensures that no 
singularity is encountered as $\lambda$ increases.

Let us compare this with the results of Sec.\,\ref{SEC:4}.  For $\lambda\gg\lambda_c$ the asymptotic form is $\Delta P|_\mathrm{3-loop}=\num{5.3E-20}i\Lambda^4 \lambda^{13}$ while for the two-loop diagram no analytic form was derived. Therefore, we will work with a fit to numerical data between $\lambda=200=14.42\lambda_c$ and $\lambda=1000=72.10\lambda_c$ which yields  $\Delta P|_\mathrm{2-loop}=\num{-5E-5}\Lambda^4\lambda^{1.4}$. Using this, the high-temperature behaviour of $f(\lambda)$ is given as 
\[
\begin{split}
f(\lambda)&=\frac{-\num{5E-5}\lambda^{1.4}}{-\num{5E-5}\lambda^{1.4}-5.3\cdot 10^{-20}i\lambda^{13}}\\
&=-\num{0.94E15}i\lambda^{-11.6}\frac{1}{1-\num{0.94E15}i\lambda^{-11.6}}
\approx  -\num{0.94E15}i\lambda^{-11.6}\,.
\end{split}
\]
It follows that the resummed three-loop contribution (with both vertices resummed) becomes
\[\label{eq:3lresummed}
\begin{split}
f^2(\lambda)\Delta P|_\mathrm{3-loop}&\approx \left(-\num{0.94E15}i\lambda^{-11.6}\right)^2\cdot5.3\cdot 10^{-20}i\Lambda^4\lambda^{13}\\
&=-\num{4.7E10}i\Lambda^4\lambda^{-10.2}
\end{split}
\]
which still is purely imaginary. Moreover, the resummed three-loop contribution is now sufficiently suppressed at high temperatures compareded to $P|_\mathrm{1-loop}\,$, as is the resummed two-loop contribution
\[\label{eq:2lresummed}
\begin{split}
f(\lambda)\Delta P|_\mathrm{2-loop}&\approx -\num{0.94E15}i\lambda^{-11.6}\left(\num{-5E-5}\Lambda^4\lambda^{1.4}\right)\\
&=\num{4.7E10}i\Lambda^4\lambda^{-10.2}.
\end{split}
\]
In fact, it is rigorously guaranteed that the leading orders in $\lambda$ of the resummed three-loop and two-loop diagrams \eqref{eq:3lresummed} and \eqref{eq:2lresummed}, respectively, cancel. This is because 
\[
f(\lambda)\approx -\frac{\Delta P|_\mathrm{2-loop}}{\Delta P|_\mathrm{3-loop}}
\]
such that
\[
\begin{split}
f^2(\lambda)\Delta P|_\mathrm{3-loop} &=\phantom{-}  \frac{(\Delta P|_\mathrm{2-loop})^2}{\Delta P|_\mathrm{3-loop}} \\
f(\lambda)\Delta P|_\mathrm{2-loop}&= -\frac{(\Delta P|_\mathrm{2-loop})^2}{\Delta P|_\mathrm{3-loop}}\,.
\end{split}
\]
Amusingly, if we pretend that \eqref{eq:foflambda} can be extended down to $\lambda_c$, we can infer from  $|\Delta P|_\mathrm{2-loop}|\gg|\Delta P|_\mathrm{3-loop}|$ that $f(\lambda)\approx1$ at low temperatures. This would mean that in the vicinity of $\lambda_c$ 
no modification of the bare 4-vertex occurs. Let us again comment on the imaginary nature of the resummed two-loop and three-loop corrections to the pressure. (The fact that their leading powers in $\lambda$ cancel does not necessarily extend to subleading powers.) The $P|_\mathrm{1-loop}$ was derived demanding thermodynamical selfconsistency at one-loop level. The impossibility to extend thermodynamical selfconsistency beyond one-loop order is 
reflected by small (and thus well controlled) imaginary and thus non-thermal corrections to the pressure which arise 
upon resummation of 2PI diagrams with dihedral symmetry.

\section{\label{SEC:6}Summary, Conclusions, and Outlook}

In the present work we have readdressed the loop expansion in the deconfining phase of SU(2) Yang-Mills 
thermodynamics. Upon analysing the constraining power imposed by 4-vertices on loop integrations in the massive sector, arising from kinematic restriction on momentum 
transfer due to the principle unresolvability of (anti)caloron centers, in terms of allowed energy-sign and scattering-channel combinations, we have found that there is an infinite class ${\cal C}$ of 2PI diagrams, exhibiting dihedral 
symmetry, with non-vanishing support of their loop integrations. The expectation 
that the loop expansion terminates at a finite 2PI loop order \cite{hofmann06} 
thus is refuted. Upon further analysis of the first member of 
class ${\cal C}$ we arrive at the interesting situation that, for temperatures not far above 
the critical temperature of the deconfining-preconfining phase transition 
\cite{Hofmann2016}, a hierarchical suppression of three-loop vs. two-loop vs. 
one-loop is observed while the modulus of the three-loop pressure 
correction rises by a high power at high temperatures, and thus is unbounded compared 
to lower loop orders. Based on the expectation that {\sl fixed} loop orders should represent small 
corrections to the a priori estimate posited by the thermal ground state and its free 
quasiparticle excitations this would represent a disaster. 

However, there definitely is a way out of this apparent dilemma. 
Namely, a formal resummation of {\sl all} diagrams in class ${\cal C}$ by virtue of the 
truncated Dyson-Schwinger equation for the vertex yields an extremly well bounded 
high-temperature dependence which is analytically connected to the 
low-temperature situation. Moreover, we show that this resummed correction 
to the free quasiparticle pressure is {\sl imaginary} at high temperatures, and 
thus, albeit strongly suppressed, not interpretable as a thermal effect. The leading (highly negative) 
powers in two-loop and 2PI three-loop corrections to the pressure, when 
subjected to resummed 4-vertices, cancel, however. There is an interesting geometric interpretation 
of (small) imaginary radiative corrections in thermodynamical quantities. Namely, the a priori estimate of the thermal 
ground state assumes that the dense packing of ball-shaped (anti)caloron centers yields a spatially homogeneous 
situation. However, there are packing voids violating this assumption, and we may attribute the implied correction 
to spatial homogeneity to the presence of turbulence-related imaginary contributions in the loop expansion.      

It will be interesting to analyse further the less symmetric five- and six-loop diagrams of 
Secs.\,\ref{5loop} and \ref{6loop} in terms of their high-temperature 
behaviour. We expect that they are much more constrained, likely so much that 
resummations no longer are required.

\section*{Acknowledgments}
We thank the Centre National de la Recherche 
Scientifique (CNRS) for the funding of RH during a one-month stay at INLN (Nice) in August-September 2016 where, among other projects, this work was conceived, and IB would like to thank ITP of the University of Heidelberg 
to fund his stay at INLN during a part of this period.

\appendix
\section{Derivation of equation \eqref{eq:3-loop-start}\label{app1}}
By rescaling loop momenta with $|\phi|$ and considering overall four-momentum conservation, one shows in a straightforward way that a bubble diagram with $n$ 4-vertices generates the following schematic contribution to the  pressure
\[\label{eq:scaling}
\frac{I_n}{(4\pi)^4\delta^{(4)}(0)} = |\phi|^{4\cdot2n}|\phi|^{-4\cdot(n-1)}|\phi|^{-2\cdot2n}\tilde I_n=|\phi|^4\tilde I_n=\left|\sqrt{\frac{\Lambda^2}{\lambda}}\right|^4\tilde I_n=\frac{\Lambda^4}{\lambda^2}\tilde I_n\,,
\]
where $I_n$ represents the loop integral arising by application of Feynman rules, and $\tilde I_n$ is the according integral in dimensionless integration variables. Recall that $\lambda=2\pi T/\Lambda$. Equation \eqref{eq:scaling} makes explicit that the mass dimension of pressure is four.

Straightforward application of the Feynman rules \cite{Hofmann2016} to the diagram in \autoref{fig:3-loop-1-b} yields (dimensionless integration variables and $m$)
\begin{multline}
\Delta P|_\mathrm{3-loop}=\frac{\Lambda^4}{96i\lambda^2}\intkern\frac{\der^4p_1\der^4p_2\der^4p_3\der^4p_4}{(2\pi)^{16}}(-ie^2)^2(2\pi)^4\delta^{(4)}(-p_1-p_4+p_3+p_2)\\
\times\zeta^{\mu_1\nu_1\rho_1\sigma_1}_{a_1b_1c_1d_1}\zeta^{\mu_2\nu_2\rho_2\sigma_2}_{a_2b_2c_2d_2}
(-2\pi)^4\delta_{c_2a_1}\delta_{c_1a_2}\delta_{d_1b_2}\delta_{d_2b_1}
\tilde D_{\rho_2\mu_1}(p_1)\tilde D_{\rho_1\mu_2}(p_3) \\
\times \tilde D_{\sigma_1\nu_2}(p_2) \tilde D_{\sigma_2\nu_1}(p_4)\delta(p^2_1-m^2)\delta(p^2_2-m^2) \delta(p^2_3-m^2)\delta(p^2_4-m^2)\\
\times n_B'(|\mathbf{p}_1|)n_B'(|\mathbf{p}_2|)n_B'(|\mathbf{p}_3|)n_B'(|\mathbf{p}_4|)\,,
\end{multline}
where $n_B'(|\mathbf{p}|)\equiv n_B(2\pi\sqrt{|\mathbf{p}|^2+m^2}/\lambda^{3/2})$ is introduced as a shorthand. 
The tensor structure $\zeta^{\mu\nu\rho\sigma}_{abcd}$ is defined as
\begin{align}
\zeta^{\mu\nu\rho\sigma}_{abcd} &\equiv \epsilon_{fab}\epsilon_{fcd}(g^{\mu\rho}g^{\nu\sigma}-g^{\mu\sigma}g^{\nu\rho})+ \\
&\qquad\qquad\epsilon_{fac}\epsilon_{fdb}(g^{\mu\sigma}g^{\rho\nu}-g^{\mu\nu}g^{\rho\sigma})+
\epsilon_{fad}\epsilon_{fbc}(g^{\mu\nu}g^{\sigma\rho}-g^{\mu\rho}g^{\sigma\nu})\,,
\end{align}
where $\epsilon_{abc}$ denotes the Levi-Civita symbol, and the propagator's tensor $\tilde D_{\mu\nu}(p)$ reads
\[
\tilde{D}_{\mu\nu}(p)=g_{\mu\nu}-\frac{p_\mu p_\nu}{m^2}\,.
\]
First, we sum over color (latin) indices, ranging from 1 to 2 (massive sector only). The contractions involving Kronecker symbols yield
\begin{multline}\label{eq:3-loop-step-2}
\frac{\Lambda^4}{96i\lambda^2}\intkern\frac{\der^4p_1\der^4p_2\der^4p_3\der^4p_4}{(2\pi)^{16}}(-ie^2)^2(2\pi)^4\delta^{(4)}(-p_1-p_4+p_3+p_2)\zeta^{\mu_1\nu_1\rho_1\sigma_1}_{abcd}\zeta^{\mu_2\nu_2\rho_2\sigma_2}_{cdab}\\
\times(-2\pi)^4\tilde D_{\rho_2\mu_1}(p_1)\tilde D_{\rho_1\mu_2}(p_3) \tilde D_{\sigma_1\nu_2}(p_2) \tilde D_{\sigma_2\nu_1}(p_4) \delta(p^2_1-m^2)\delta(p^2_2-m^2)\\
\times \delta(p^2_3-m^2)\delta(p^2_4-m^2)n_B'(|\mathbf{p}_1|)n_B'(|\mathbf{p}_2|)n_B'(|\mathbf{p}_3|)n_B'(|\mathbf{p}_4|) 
\end{multline}
with a relabelling $a_1, b_1, c_1, d_1\to a,b,c,d$. 
Because of the restriction to the massive sector, we have $\epsilon_{fmn}=\epsilon_{3mn}=\epsilon_{mn}$ for $m,n=1,2$, and one can write
\[\label{eq:contractedzetas}
\begin{split}
&\zeta^{\mu_1\nu_1\rho_1\sigma_1}_{abcd}\zeta^{\mu_2\nu_2\rho_2\sigma_2}_{cdab}\\
=&\left[\epsi{ab}\epsi{cd}G^{A}_{1}+\epsi{ac}\epsi{db}G^{B}_{1}
+\epsi{ad}\epsi{bc}G^{C}_{1} \right] \cdot\left[\epsi{cd}\epsi{ab}G^{A}_{2}+\epsi{ca}\epsi{bd}G^{B}_{2}+\epsi{cb}\epsi{da}G^{C}_{2}\right] \\
=& 4\left(G^{A}_{1}G^{A}_{2}+G^{B}_{1}G^{B}_{2}+G^{C}_{1}G^{C}_{2}\right)\\
&\qquad-2\left(G^{A}_{1}G^{B}_{2}+G^{A}_{1}G^{C}_{2}+G^{B}_{1}G^{A}_{2}+G^{B}_{1}G^{C}_{2}+G^{C}_{1}G^{A}_{2}+G^{C}_{1}G^{B}_{2}\right),
\end{split}
\]
where the superindices $A$, $B$, and $C$ encode the Lorentz structure as
\[
\begin{split}
G^{A}_{i}&\equiv g^{\mu_i\rho_i}g^{\nu_i\sigma_i}-g^{\mu_i\sigma_i}g^{\nu_i\rho_i} \\
G^{B}_{i}&\equiv g^{\mu_i\sigma_i}g^{\rho_i\nu_i}-g^{\mu_i\nu_i}g^{\rho_i\sigma_i} \\
G^{C}_{i}&\equiv g^{\mu_i\nu_i}g^{\sigma_i\rho_i}-g^{\mu_i\rho_i}g^{\sigma_i\nu_i}\,.
\end{split}
\]
Relabelling $\mu_1,\nu_1,\rho_1,\sigma_1\rightarrow\mu,\nu,\rho,\sigma$ and $\mu_2,\nu_2,\rho_2,\sigma_2,\rightarrow\bar\mu,\bar\nu,\bar\rho,\bar\sigma$, \eqref{eq:contractedzetas} evaluates to
\begin{multline}
12\left(g^{\mu\rho}g^{\nu\sigma}g^{\bar\mu \bar\rho}g^{\bar\nu\bar\sigma}
+g^{\mu\sigma}g^{\nu\rho}g^{\bar\mu \bar\sigma}g^{\bar\nu\bar\rho}
+g^{\mu\nu}g^{\rho\sigma}g^{\bar\mu \bar\nu}g^{\bar\rho\bar\sigma}\right)\\
-6\left(g^{\mu\rho}g^{\nu\sigma}g^{\bar\mu \bar\sigma}g^{\bar\nu\bar\rho}
+g^{\mu \sigma }g^{\nu \rho }g^{\bar\mu \bar\rho}g^{\bar\nu\bar\sigma}
+g^{\mu \sigma }g^{\nu \rho }g^{\bar\mu \bar\nu}g^{\bar\rho\bar\sigma}	\right.\\\left.
+g^{\mu \nu }g^{\rho \sigma }g^{\bar\mu \bar\sigma}g^{\bar\nu\bar\rho}
+g^{\mu \nu }g^{\rho \sigma }g^{\bar\mu \bar\rho}g^{\bar\nu\bar\sigma}
+g^{\mu \rho }g^{\nu \sigma }g^{\bar\mu \bar\nu}g^{\bar\rho\bar\sigma}\right)\,.
\end{multline}
The entire integral with contractions reduced to scalar products, 
upon execution of the $p_4$ integration in \eqref{eq:3-loop-step-2} by virtue of the delta function $\delta^{(4)}$, and exploiting $p_i^2=m^2$ due to delta functions $\delta$, is computed to be
\[\label{eq:3-loop-full-integral}
\begin{split}
&\Delta P|_\mathrm{3-loop}\\
=&i\frac{\Lambda^4}{96\lambda^2}e^4(2\pi)^4\intkern\frac{\der^4p_1\der^4p_2\der^4p_3}{(2\pi)^{12}}\left\{144\vphantom{\frac1{m^4}} \right.\\
&-12\frac1{m^4}\left[(p_1p_2)^2+(p_1p_3)^2+(p_1p_4)^2+(p_2p_3)^2+(p_2p_4)^2+(p_3p_4)^2 \right] \\
&+36\frac1{m^6}\left[(p_1p_2)(p_1p_3)(p_2p_3)+(p_1p_2)(p_1p_4)(p_2p_4)\right.\\
&\phantom{+12\frac1{m^4}}\left.+(p_1p_3)(p_1p_4)(p_3p_4)+(p_2p_3)(p_2p_4)(p_3p_4)\right]\\
&+12\frac1{m^8}\left[(p_1p_2)^2(p_3p_4)^2+(p_1p_3)^2(p_2p_4)^2+(p_1p_4)^2(p_2p_3)^2\right.\\
&\phantom{+12\frac1{m^4}} - (p_1p_2)(p_1p_3)(p_2p_4)(p_3p_4)-(p_1p_2)(p_1p_4)(p_2p_3)(p_3p_4)\\
&\left.\phantom{+12\frac1{m^4}}\left.\vphantom{(p_1)^2}  - (p_1p_3)(p_1p_4)(p_2p_3)(p_2p_4)\right]\right\}\\
&\quad\times\delta(p^2_1-m^2)\delta(p^2_2-m^2)\delta(p^2_3-m^2)\delta(p^2_4-m^2)\\
&\quad\times n_B'(|\mathbf{p}_1|)n_B'(|\mathbf{p}_2|)n_B'(|\mathbf{p}_3|)n_B'(|\mathbf{p}_4|)\,,
\end{split}
\]
where $p_4=p_2+p_3-p_1$. The polynomial in braced brackets is denoted by $P(\{p_i\})$ in \eqref{eq:polynomial}.

In a next step the delta functions $\delta$ are integrated away, introducing a sum over 16 possible sign combinations which can be reduced to 8 inequivalent combinations: We make the convention that the sign of $p^0_1$ is plus. Vertex constraints restrict these configurations further, see \eqref{eq:3-constraint-s}, \eqref{eq:3-constraint-t}, and \eqref{eq:3-constraint-u}.
The ``diagonal'' channels $ss$, $tt$, and $uu$ give equal results since the integrand is symmetric under permutations of $p_1$, $p_2$, $p_3$ and $p_4$ and under inversion $p_i\rightarrow-p_i$. The same is true for the ``off-diagonal'' channels $st$, $su$, and $tu$. Considering that each of the two vertices decomposes into equally weighted $s$, $u$, and $t$ contributions, there are nine copies of the integral $\Delta P|_\mathrm{3-loop}$, each weighted by $1/9$. The diagonal channels add up to $1/3\Delta P|_{\mathrm{\mathrm{3-loop},ss}}$ while the off-diagonal channels add up to $2/3\Delta P|_{\mathrm{\mathrm{3-loop},st}}$. Namely, 
recalling the results of \autoref{sec:3-loop-signs}, one notes that off-diagonal channels reduce to one allowed configuration
\[
p^0_1>0\,,\quad p^0_2>0\,,\quad p^0_3<0\,,\quad p^0_4<0 \quad \text{($st$-channel)}
\]
and that the diagonal channels admit two configurations,
\[
\begin{split}
p^0_1&>0\,,\quad p^0_2>0\,,\quad p^0_3<0\,,\quad p^0_4<0 \quad \text{or} \\
p^0_1&>0\,,\quad p^0_2<0\,,\quad p^0_3>0\,,\quad p^0_4<0 \quad \text{($ss$-channel)}\,.
\end{split}
\]
With the following parametrisation $\mathbf{p}_i\equiv r_i(\sin\theta_i\cos\varphi_i,\sin\theta_i\sin\varphi_i,\cos\theta_i)^T$ for $i=1,2,3$ and appealing to overall rotational symmetry the integrations over angles $\theta_2$, $\varphi_2$, and $\varphi_3$ can be performed trivially.
Finally, one arrives at
\[
\begin{split}\label{eq:3-loop-start-app}
\Delta P|_\mathrm{3-loop}=&i\frac{\Lambda^4}{96\lambda^2}e^4\frac1{(2\pi)^8}\sum_\mathrm{signs}\intkern\der^3p_1\der^3p_2\der^3p_3P(\{p_i\})\\
&\times\delta\left(p^0_4\mp\sqrt{r_4^2+m^2}\right)\frac{n_B'(r_1)n_B'(r_2)n_B'(r_3)n_B'(r_4)}{8|p^0_1p^0_2p^0_3p^0_4|}\\
=&i\frac{\Lambda^4}{96\lambda^2}e^4\frac1{(2\pi)^8}\sum_\mathrm{signs}
\left(\intkern\der\Omega_1\der r_1\cdot4\pi\intkern\der r_2\cdot2\pi\int^{\pi}_0\der\theta_3\intkern\der r_3\right)\\
&\times r^2_1r^2_2r^2_3\sin\theta_3P(\{p_i\})\delta\left(p^0_4\mp\sqrt{r_4^2+m^2}\right)\frac{n_B'(r_1)n_B'(r_2)n_B'(r_3)n_B'(r_4)}{8|p^0_1p^0_2p^0_3p^0_4|}\\
=&i\frac{\Lambda^4}{48\lambda^2}e^4\frac1{(2\pi)^6}\sum_{\mathrm{signs}}\intkern\der\theta_1\der\varphi_1\der r_1\der r_2\der\theta_3\sum_{\{r_3\}}r^2_1r^2_2r^2_3\sin\theta_1\sin\theta_3\\
&\times P(\{p_i\})\frac{n_B'(r_1)n_B'(r_2)n_B'(r_3)n_B'(r_4)}{8|p^0_1p^0_2p^0_3p^0_4|},
\end{split}
\] 
where $|p^0_i|=\sqrt{r_i^2+m^2}$, and in the last step the $r_3$ integration is executed by summing over solutions in $r_3$ of
\begin{multline}
\label{eq:delta4}
\sgn(p^0_2)\sqrt{r_2^2+m^2}+\sgn(p^0_3)\sqrt{r^2_3+m^2}-\sqrt{r_1^2+m^2}=
\sgn(p^0_4)\left[r^2_1+r^2_2+r^2_3 \right.\\\left. - 2 r_1r_2\cos\theta_1  -r_1r_3(\sin\varphi_1\sin\theta_1\sin\theta_3+\cos\theta_1\cos\theta_3)+r_2r_3\cos\theta_3 + m^2\right]^{1/2}\,.
\end{multline}

\par
\bigskip

\bibliographystyle{hunsrt}
\bibliography{RadCorBib}{}

\begin{thebibliography}{10}

\bibitem{Hofmann2016}
R.~Hofmann.
\newblock {\em The thermodynamics of quantum {Yang-Mills} theory}.
\newblock World Scientific Publishing Company, second edition, 2016.

\bibitem{thermalPT1}
E.V. Shuryak.
\newblock {\em Sov.Phys.JETP}, 47:212, 1978.

\bibitem{thermalPT2}
E.~Braten and R.~Pisarski.
\newblock {\em Phys.Rev.Lett.}, 64:1338, 1990.

\bibitem{thermalPT3}
E.~Braten and R.~Pisarski.
\newblock {\em Nucl.Phys.B}, 337:569, 1990.

\bibitem{thermalPT4}
J.~Frenkel and J.C. Taylor.
\newblock {\em Nucl.Phys.B}, 334:199, 1990.

\bibitem{thermalPT5}
P.~Arnold and Ch. Zhai.
\newblock {\em Phys.Rev.D}, 50:7603, 1994.

\bibitem{thermalPT6}
Ch. Zhai and B.~Kastening.
\newblock {\em Phys.Rev.D}, 52:7232, 1994.

\bibitem{thermalPT7}
E.~Braten and A.~Nieto.
\newblock {\em Phys.Rev.D}, 53:3421, 1996.

\bibitem{Linde1980}
A.D. Linde.
\newblock {\em Phys.Lett.B}, 96(3):289--292, 1980.

\bibitem{thermalPT8}
K.~Kajantie, M.~Laine, K.~Rummukainen, and Y.~Schroeder.
\newblock {\em Phys.Rev.D}, 67:105008, 2003, arXiv:hep-ph/0211321.

\bibitem{BGHwip}
I.~Bischer, T.~Grandou, and R.~Hofmann.
\newblock work in progress.
\newblock 2017.

\bibitem{HS1977}
B.~J. Harrington and H.~K. Shepard.
\newblock {\em Phys.Rev.D}, 17:2122--2125, Apr 1978.

\bibitem{Entropy2016}
R.~Hofmann.
\newblock {\em Entropy}, 18(19)(1):310, 2016, arXiv:1604.05136.

\bibitem{Evans}
T.~S. Evans.
\newblock What is being calculated in thermal field theory.
\newblock {\em Proc. of {``Particle physics and cosmology'', Lake Louise}},
  page 343, 1994.

\bibitem{Brodsky}
S.~J. Brodsky and P.~Hoyer.
\newblock {\em Phys.Rev.D}, 83:045026, 2010, arXiv:1009.2313.

\bibitem{SGH2007}
M.~Schwarz, R.~Hofmann, and F.~Giacosa.
\newblock {\em Int.J.Mod.Phys.A}, 22:1213--1238, 2007, arXiv:hep-th/0603078.

\bibitem{KavianiHofmann}
D.~Kaviani and R.~Hofmann.
\newblock {\em Mod.Phys.Lett.A}, 22:2343--2352, 2007, arXiv:0704.3326.

\bibitem{hofmann-krasowski}
N.~Krasowski and R.~Hofmann.
\newblock {\em Ann.Phys.}, 347:287--308, 2014, arXiv:1301.4716.

\bibitem{kleinert}
H.~Kleinert, A.~Pelster, B.~Kastening, and M.~Bachmann.
\newblock {\em Phys.Rev.E}, 62:1537--1559, 2000, arXiv:hep-th/9907168.

\bibitem{BischerMA2017}
I.~Bischer.
\newblock {\em Master thesis}.
\newblock Universit\"at Heidelberg, first edition, 2017.

\bibitem{herbst-hofmann-rohrer}
U.~Herbst, R.~Hofmann, and J.~Rohrer.
\newblock {\em Acta Phys.Polon.B}, 36:881--904, 2004, arXiv:hep-th/0410187.

\bibitem{hofmann06}
R.~Hofmann.
\newblock {\em Braz.J.Phys.}, 42:110--119, 2006, arXiv:hep-th/0609033.

\bibitem{KellerRH2008spatialWilsonloop}
J.~Ludescher~{et al.}
\newblock {\em Ann.d.Phys.}, 19:102--120, 2010, arXiv:0812.1858.

\bibitem{swanson}
E.~S. Swanson.
\newblock {\em AIP Conf.Proc.}, 1296:75--121, 2010, arXiv:1008.4337.

\end{thebibliography}

\end{document}